\documentclass[12pt,preprint]{aastex}
\usepackage{float}
\usepackage{arydshln}

\newcommand{\Msun}{\hbox{M$_{\odot}$}}

\newcommand{\Teff}{\mbox{$T_{\rm eff}$}}

\newcommand{\Lya}{Lyman~$\alpha$}
\newcommand{\HST}{{\sl HST}}

\newcommand{\rosat}{{\sl ROSAT}}

\newcommand{\galex}{{\sl GALEX}}

\slugcomment{Accepted to the Astronomical Journal}

\shorttitle{HAZMAT I: The Evolution of Far-UV and Near-UV Emission from Early M Stars
}
\shortauthors{Shkolnik \& Barman}

\begin{document}


\title{HAZMAT I: The Evolution of Far-UV and Near-UV Emission from Early M Stars
\altaffilmark{1}\\}


\author{Evgenya~L.~Shkolnik}
\affil{Lowell Observatory, 1400 West Mars Hill Road, Flagstaff, AZ, 86001, USA}
\email{shkolnik@lowell.edu}

\and

\author{Travis S. Barman}
\affil{Department of Planetary Sciences and Lunar and Planetary Laboratory
University of Arizona, Tucson AZ, 85721, USA}
\email{barman@lpl.arizona.edu}

\altaffiltext{1}{Based on observations made with the NASA Galaxy Evolution Explorer.
\galex\ was operated for NASA by the California Institute of Technology under NASA contract NAS5-98034.}

\begin{abstract}

 The spectral energy distribution, variability and evolution of the high-energy radiation from an M dwarf planet host  is crucial in understanding the planet's atmospheric evolution and habitability and in interpreting the planet's spectrum. The star's extreme-UV (EUV), far-UV (FUV) and near-UV (NUV) emission can chemically modify, ionize, and erode the atmosphere over time. This makes determining the lifetime exposure of such planets to stellar UV radiation critical for both the evolution of a planet's atmosphere  and our potential to characterize it.   Using the early M star members of nearby young moving groups (YMGs), which sample critical ages in planet formation and evolution, we measure the \galex\ NUV and FUV flux as a function of age.  The median UV flux remains at a ``saturated'' level for a few hundred million years,  analogous to that observed for X-ray emission. By the age of the Hyades Cluster (650 Myr), we measure a drop in UV flux  by a factor of 2--3 followed by a steep drop from old (several Gyrs) field stars. This decline in activity  beyond 300 Myr follows  roughly $t^{-1}$.  Despite this clear evolution, there remains a wide range of 1--2 orders of magnitude in observed emission levels at every age.  
These UV data supply the much-needed constraints to M dwarf upper-atmosphere models, which will  provide empirically-motivated EUV predictions and more accurate age-dependent UV spectra as inputs to planetary photochemical models.

\end{abstract}



\section{Introduction}\label{intro}

The conditions inside the circumstellar disks of M dwarfs provide a favorable environment for the formation of low-mass planets close to the star \citep{wu13}, as indicated by the $\sim$50\% small planet occurrence rate around M dwarfs in the habitable zone (HZ; $\approx$0.1--0.4 AU; \citealt{bonf13,kopp13,dres14}).  This implies that most of the planets in our galaxy, including those in the HZ, orbit M dwarfs as these low-mass stars make up 75\% of the total stellar population (\citealt{boch08}). In fact, M dwarfs are proposed \citep{tart07,scal07} to be the most sought-after candidate planet-hosts as they are the most amenable to follow-up observations, including the first potentially habitable planets to be spectroscopically characterized, probably by $JWST$ transit transmission observations (\citealt{demi09}).   In the next few years, many more planets will be identified around bright M stars by both ground- and space-based instruments including TESS, CARMENES, SPIROU and NGTS (\citealt{rick14,quir12,delf13,whea14}, respectively), for which the incident stellar high-energy spectrum will be critical to understanding the planets' atmospheres. 
 
The EUV and FUV spectrum of the star dominates atmospheric photochemistry of planets by affecting composition, ionization, and stability 
(e.g., \citealt{kast93,lich10,segu10}), including the photodissociation of  molecules important to surface habitability, e.g.~H$_2$O, CH$_4$, and CO$_2$. The UV increases the generation of surface-shielding hazes in reducing atmospheres (\citealt{zerk12}) and ozone (O$_3$) in oxidizing atmospheres (\citealt{segu03,segu05}), both of which can strongly affect the observed spectrum. Recently, \cite{robi14} showed that Titan's high altitude haze affects its transit spectrum and severely limits the atmospheric depths which can probed by such data.   Observations of exoplanet transit transmission spectroscopy (\citealt{bean11,krei14}) are showing that hazes in planets around M dwarfs might in fact be quite common, probably generated by the star's UV.   The ratio of the FUV to NUV flux  can increase the detectability of biologically generated gases (\citealt{segu05}), but may also lead to the formation of abiotic oxygen and ozone (\citealt{tian14,doma14}) producing a false-positive biosignature for oxygenic photosynthesis. And the NUV flux can photodissociate diagnostic molecules such as sulfur dioxide (SO$_2$) and ammonia (NH$_3$).

The EUV stellar radiation (100--900\AA) can be particularly damaging as it photoionizes, heats and inflates a planet's upper-atmosphere making it vulnerable to massloss \citep{kosk10}. Two aeronomical studies have explored the potential for massloss by applying a wide range of EUV emission. \cite{lamm07} conclude that the atmosphere
of an unmagnetized telluric planet can be completely eroded in its first Gyr by high M dwarf activity levels and corresponding coronal mass ejections. On
the other hand, \cite{tian09} find that the atmosphere of a super-Earth is stable even around very active M dwarfs.  

\begin{figure}[tbp]
	\includegraphics[width=5.in,angle=270]{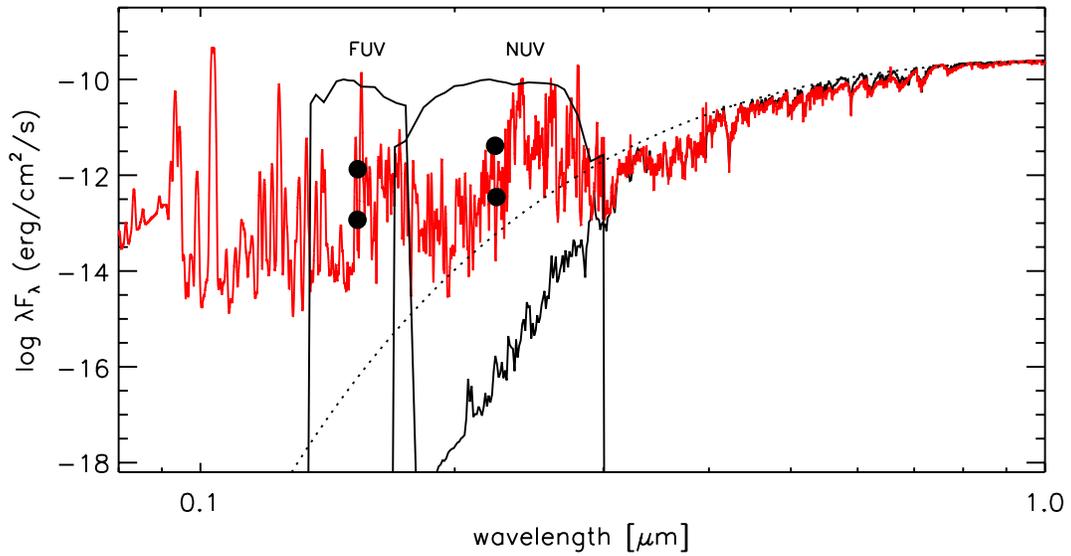}
		\caption{Model spectra for a 50-Myr old star with \Teff=3500 K:  photosphere only model (solid black curve), the photosphere + upper-atmosphere model  (red curve),  and a 3500-K blackbody  (black dashed curve).  The data points are \galex\ fluxes for a young (higher points) and old star (lower points) with the same \Teff.\label{model_spec}}
	\end{figure}

Direct observations of the EUV spectral range is currently impossible since the completion of the EUVE space mission and its obscuration due to interstellar neutral hydrogen.  Estimating EUV fluxes from existing M dwarf model atmospheres is also not feasible because current models lack any prescription for
the lowest density regions of the upper-atmosphere, namely the chromosphere,
transition region and corona, causing severely under-predicted emission at all UV wavelengths (e.g.~\citealt{woit11} and Figure~\ref{model_spec}). 

Two empirical options remain: extrapolate to EUV wavelengths from either X-ray or UV observations. \cite{leca07} and \cite{sanz11} estimate EUV fluxes in known, old, solar-type
planet hosts using coronal models fits to X-ray spectra. The latter paper noted that the
contribution to the EUV flux from the transition region is unknown because of the lack of FUV spectra. Since both the FUV and the EUV spectral ranges are filled by emission lines formed at upper-transition region and lower-coronal temperatures (e.g.~as show for the Sun by \citealt{kret09}), there is also promise to approaching the EUV predictions from the FUV \citep{lins14}. Interpolating between X-ray and FUV data for a given star is also an option, but as discussed in Section~\ref{correlations}, it is problematic with the currently available data sets.

Photochemical models calculated for the atmospheres of all exoplanets, from Earths to Jupiters (e.g., \citealt{segu10,line10,kalt11,hu12,kopp12,mose13}) require realistic input stellar fluxes, and are at the moment limited to using solar data in most cases (e.g.~\citealt{font09}). 
For M dwarfs, there exist very few UV spectra to use. \cite{walk08} collected very low-resolution Hubble Space Telescope (\HST) NUV data of 33 M dwarfs.  More recently \cite{fran13} secured \HST\ high-resolution FUV and NUV spectra of six old M dwarf planet hosts with a wide range of spectral types (M1--M6).  These are providing valuable initial inputs to planetary photochemical models \citep{migu14,tian14}, 
however, the full history of a planet's UV exposure is impossible to predict from a single observation. It is critical to study stars spanning a wide range of ages with many stars at each age to help mitigate the effects of flaring from a single star and provide a more accurate mean level and range of UV activity for M stars at a given mass and age.

It is well-known that all types of stars are more active in their youth, with a progressive decline with age.  The rate of decline in stellar X-ray emission was shown to vary with stellar mass, with M dwarfs remaining X-ray active (and rotating faster) much longer than FGK stars \citep{pizz03,prei05,sels07}.
The Galaxy Evolution Explorer (\galex) provides a new data set with which to study the broadband FUV and NUV emission from many more stars, and at much greater distances ($\gtrsim$100 pc) than previously possible \citep{find10,shko11a}, including 65\% of the known planet hosts (\citealt{shko13}). Here, we detail the UV evolution of early Ms, from 10's to 100's to 1000's of Myrs, sampling critical ages in planet formation and evolution (e.g., \citealt{mand07}), extending the work of \cite{find11} for more massive stars to M dwarfs. 
This work represents the first results of the HAZMAT (HAbitable Zones and M dwarf Activity across Time) program, providing the empirical guidance needed to build new M dwarf upper-atmosphere models (Peacock, Barman \& Shkolnik, in preparation), to characterize of the full-UV spectrum, including the EUV and \Lya, for the stars that are the most common planet hosts. These models will provide a grid of input spectra to planetary atmospheric photochemical models to study the impact of M dwarf UV evolution on planetary atmospheres, including constraining planet atmospheric evaporation which is completely dependent upon the stellar EUV fluxes.

The target stars in this study consist of the low-mass members of the nearby young moving groups (YMGs), which provide the most accurate stellar ages available for dispersed M stars. The YMGs are TW Hydra at 10 Myr, $\beta$~Pic at 12 Myr,\footnote{\cite{bink13} places the $\beta$ Pic YMG at 21$\pm$4 Myrs.} Tuc-Hor at 40 Myr, AB Dor at 100 Myr, Ursa Major at 300 Myr, and the Hyades cluster at 650 Myr. Each star has been shown to be kinematically linked (using 3-D space velocities) to one of these YMGs (e.g., \citealt{zuck04,torr08,shko11a,shko12,krau14}), and also exhibits independent youth indicators such as elevated stellar activity levels, low gravity, and possibly lithium absorption (e.g.~\citealt{shko09b}).  We complement the study with the old population of M stars within 10 pc, which has an average age of $\sim$5 Gyr.


\section{\galex\ NUV and FUV Photometry}

The \galex\ satellite was launched on April 28, 2003 and imaged
approximately 3/4 of the sky simultaneously in two UV bands: FUV 1350--1750 \AA\ and NUV 1750--2750 \AA.
For stars hotter than about 5250 K, the flux in the \galex\ bandpasses is made up predominantly from continuum emission \citep{smit10} with additional flux provided by strong  emission lines (C IV, C II, Si IV, He II) originating from the upper-atmosphere.  Cooler stars have FUV and NUV fluxes strongly dominated by stellar activity (e.g.~\citealt{robi05,wels06,paga09a}), making \galex\ an excellent tool with which to study stellar activity in low-mass stars lying within $\sim$150 pc (e.g.~\citealt{find10,shko11a}). 
 The \galex\ FUV bandpass does not include the chromospheric \Lya\ line (1216 \AA), which \cite{lins14} measure to be as bright as the entire 1200--3200~\AA\ spectrum of M dwarfs \citep{fran13}, but direct observations of the  \Lya\ emission are very difficult due to interstellar hydrogen absorption and geocoronal emission. \cite{lins13} showed that intrinsic \Lya\ correlates with other emission lines, including C~IV, which contributes $\sim$50\% of the \galex\ FUV flux (\citealt{robi05,wels07}).

In addition to a medium and a deep imaging survey (MIS, DIS), covering 1000 and 100 square
degrees, respectively, the \galex\ mission has produced an All-sky Imaging Survey (AIS), all of which is archived at
the Barbara A. Mikulski Archive for Space Telescopes (MAST). The angular
resolutions are 6.5\arcsec\ and 5\arcsec\ in the FUV and NUV, respectively, across a 1.25$^{\circ}$ field of view. 
The full description of the instrumental
performance is presented by \cite{morr05}.\footnote{One can query the \galex\ archive through either
CasJobs (http://mastweb.stsci.edu/gcasjobs/) or the web tool GalexView (http://galex.stsci.edu/galexview/).} The
fluxes and magnitudes averaged over the entire exposure were produced by the standard \galex\ Data
Analysis Pipeline (ver.~4.0) operated at the Caltech Science Operations Center \citep{morr07}. The data presented in this
paper made use of the sixth data release (GR6/7), which includes the three surveys plus publicly available data from Guest Investigator (GII) programs.

The \galex\ pipeline performs aperture photometry using several sizes. We chose the ``aper\_7'' which has a radius of 17.3\arcsec. This relatively large aperture requires the least aperture correction (0.04 mags in both the in NUV and FUV bandpasses)\footnote{See Table 1 of http://www.galex.caltech.edu/researcher/techdoc-ch5.html. Note \cite{morr07} quote a required aperture correction of 0.07 mags.  Either way the effect is very small compared to the uncertainties and the large differences in flux between targets.} and encompasses the full range of possible PSFs, even near the edges of the images where the PSF is elongated. In order to exclude the severest edge effects, we limit our detections to those within 0.59$^\circ$ from the center of the 1.25$^\circ$-wide image.

\begin{figure}[tbp]
	\plotone{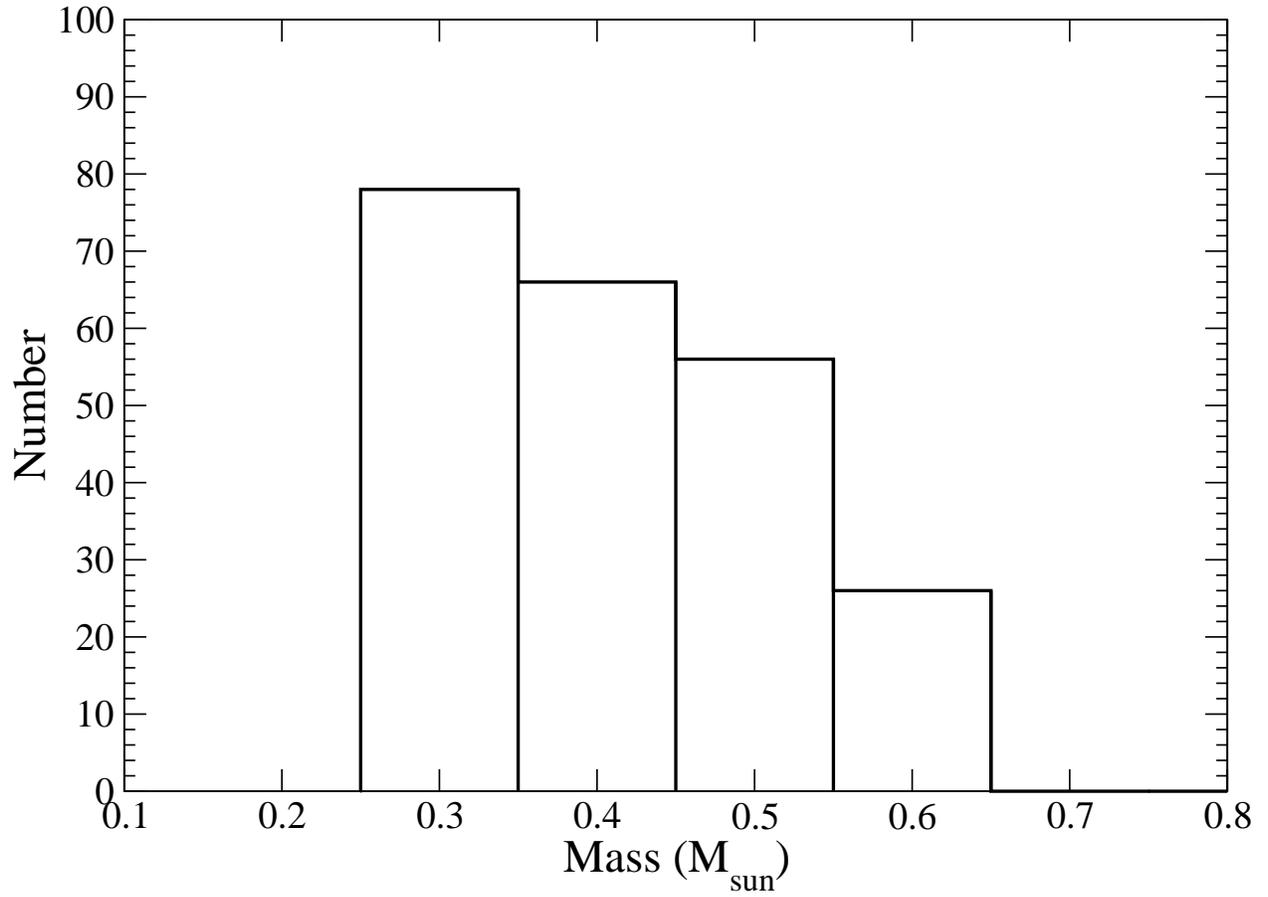}
		\caption{Stellar mass distribution of the sample using literature SpTs and \cite{bara98} models for a given stellar age. \label{mass_hist}}
	\end{figure}

We cross-correlated the published early-M YMG members as of April 2014 and old field stars with masses ranging from $\approx$0.3--0.65\Msun\ with the \galex\ archive using a 12\arcsec\ search radius (Figure~\ref{mass_hist}) resulting in 215 observed stars. We exclude lower mass targets at this time because there are very few known in YMGs, and since such fully-convective stars have been shown to exhibit much more erratic flare activity (e.g.~\citealt{rein09b}),  it would be difficult to make statistical conclusions about the evolution of their stellar activity with so few. 

\cite{krau14} showed that many of these young stars originally selected for their youth using high UV emission (e.g.~\citealt{shko11a,rodr11}) are not biased towards only the high-activity stars, but rather the vast majority of all YMG members are indeed very active.  For the old sample of stars within 10 pc that have very high proper motions, we conducted the search with proper-motion-corrected coordinates using the dates of the \galex\ images.  Fluxes are calculated from the reported flux densities using the effective wavelengths of 2267\AA\ and 1516\AA\ for the NUV and FUV bandpasses, respectively.\footnote{These values are taken from Table 1.1 of\\ http://galexgi.gsfc.nasa.gov/docs/galex/Documents/ERO\_data\_description\_2.htm.  \cite{morr07} report slightly different values for effective wavelengths: 2315.7~\AA\ and 1538.6~\AA.} Given that our sample is mostly within 50 pc, we do not include the effects of UV extinction as \cite{find11} showed that the effects are insignificant even out to 250 pc.

\begin{figure}[tbp]
	\plotone{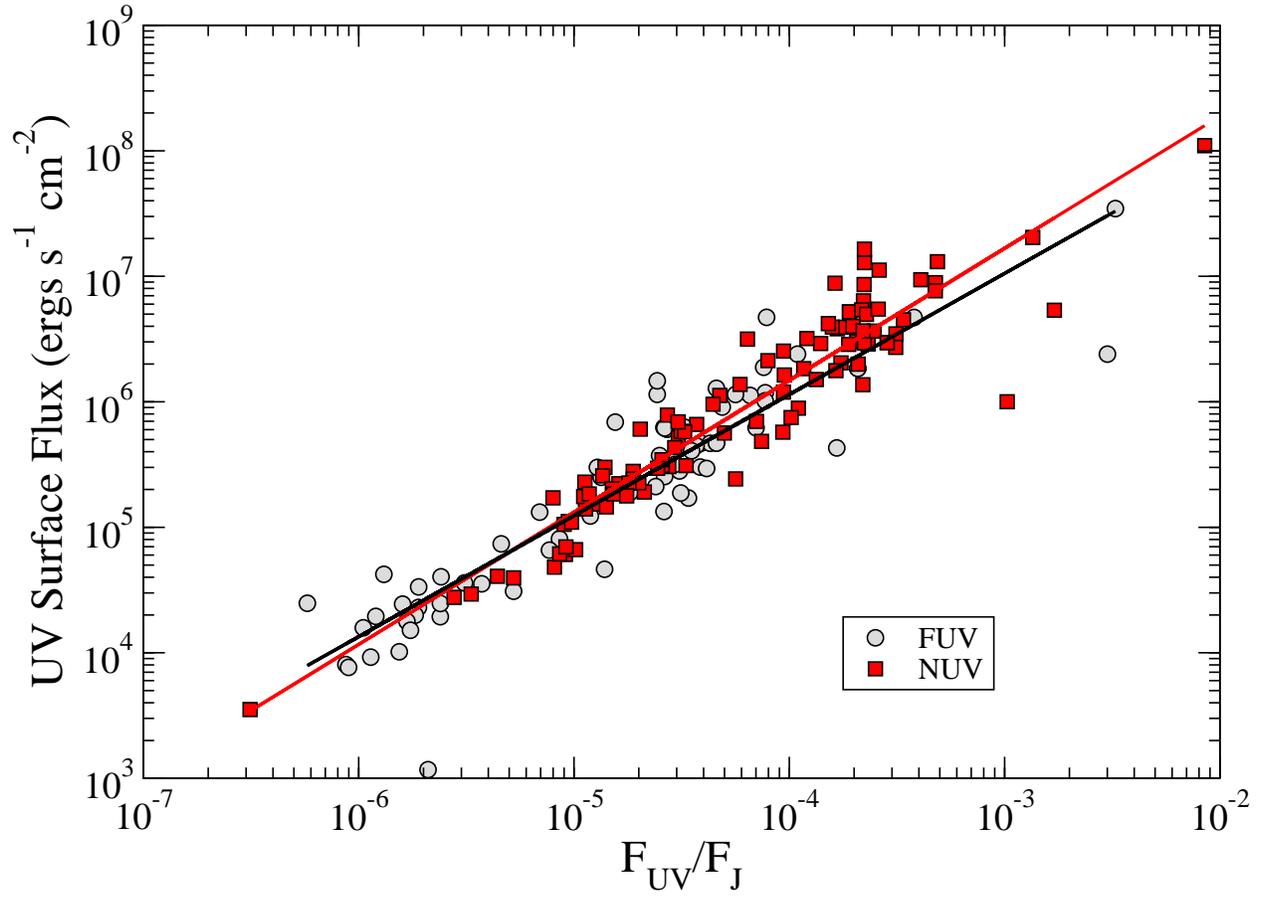}
		\caption{FUV and NUV fractional flux densities plotted against the surface flux for those stars with trigonometric parallaxes. \label{uv_fractional_surface_flux}}
	\end{figure}

Since not all of the stars have trigonometric parallaxes, nor are published bolometric corrections for M dwarfs precise, we analyze the UV fractional flux densities relative to the 2MASS J magnitude, i.e.~$F_{FUV}/F_J$ and $F_{NUV}/F_J$.\footnote{These flux densities are calculated using Janskys.  In order to convert these ratios to quantities based in flux units of erg~s$^{-1}$~cm$^{-2}$, multiplicative factors of 10.98 and 13.41 can be applied to the $F_{FUV}/F_J$ and $F_{NUV}/F_J$ values, respectively.}  A comparison of fractional flux densities to stellar surface flux for those stars with parallaxes is shown in Figure~\ref{uv_fractional_surface_flux}. 

\begin{figure}[tbp]
		\plotone{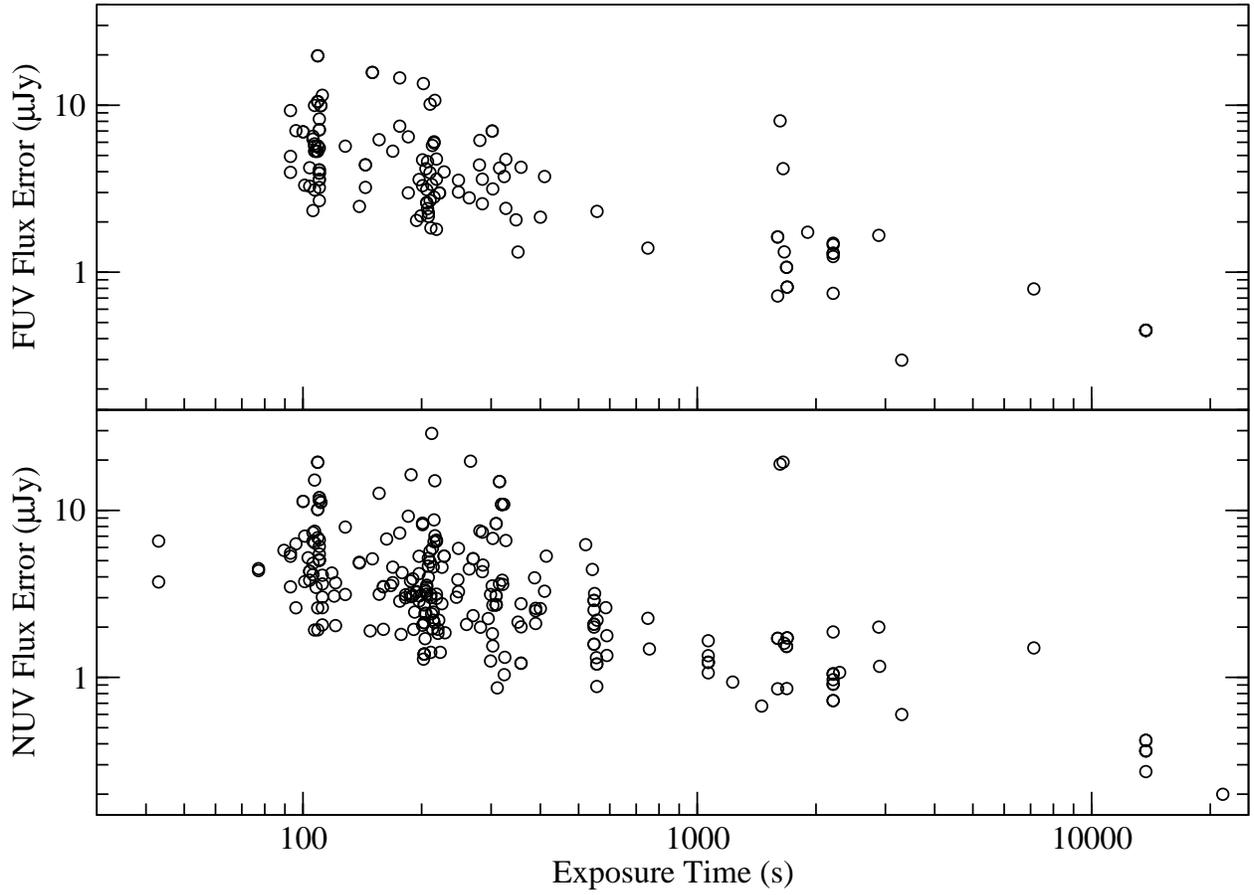}
			\caption{Reported flux errors for the FUV (top) and NUV (bottom) detections. \label{flux_err}}
		\end{figure}

Table~\ref{table_targets} lists all of our targets observed by \galex, their coordinates, published spectral types (SpTs), 2MASS J band magnitudes, and UV flux densities.\footnote{We independently measured the FUV and NUV fluxes of all the stars, including the old stars that overlap with those in \cite{stel13}. We noticed an error in \cite{stel13}'s values which they corrected in \cite{stel14}.} In the NUV, 95\% the stars observed by \galex\ were detected. In the FUV,  72\% of Tuc-Hor members,  35\% of the Hyades members, and 52\% of the old stars were detected. For those not detected, we calculated 1-$\sigma$ upper limits by determining the median flux error for a given exposure time for those stars that were detected.  These are shown in Figure~\ref{flux_err}. We identified the targets with known stellar companions within 17.3\arcsec, which may increase the observed flux either by having an active secondary or through the tidal spin-up of primary. It is evident that the TW Hydra Association (TWA; 10 Myr) is so far the best-surveyed YMG for companions (64\% of our TWA sample are known binaries), and given that our knowledge of the other YMGs is not nearly as complete, we have chosen to include the known binaries in our analysis, with the assumption that the binary fraction in other moving groups is likely to be comparable to TWA's. Including the known binaries does not significantly affect the final conclusions of this paper.


\section{Evolution of the Photospheric UV Emission}\label{photospheric_flux}

\begin{figure}[tbp]
					\plotone{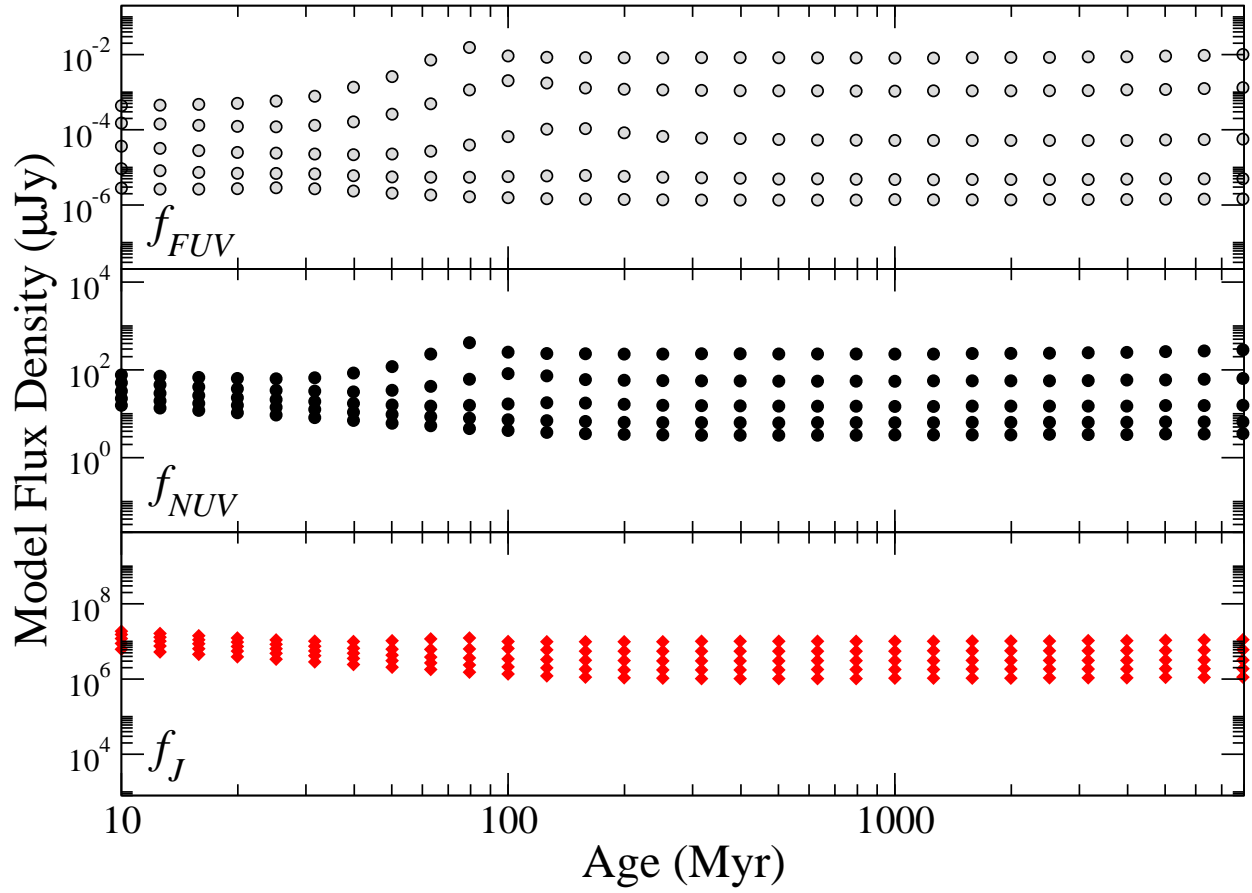}
						\caption{The photospheric FUV, NUV and J band flux density as a function of age for masses M$_{*}$=0.3, 0.4, 0.5, 0.6 and 0.7 \Msun\ (low to high). \newline \label{age_model_fluxes}}
				\end{figure}

Commonly used M dwarf photospheric models have temperature structures that do not include rises characteristic of chromospheres, transition regions, or coronae and, therefore, the term ``photospheric'' (or ``photosphere-only'') is used here when referring to fluxes from such models. 
In order to test the evolution of the photosphere and its corresponding effects on the observed changes in UV flux with time, we calculated the photospheric flux using the PHOENIX stellar atmosphere models \citep{haus97,shor05} convolved with the relevant NUV, FUV and J band normalized transmission curves. 
Figure~\ref{age_model_fluxes} shows the evolution of the photospheric fractional flux density  for a range of stellar masses (0.3, 0.4, 0.5, 0.6, and 0.7 \Msun). 

\begin{figure}[tbp]
			\plottwo{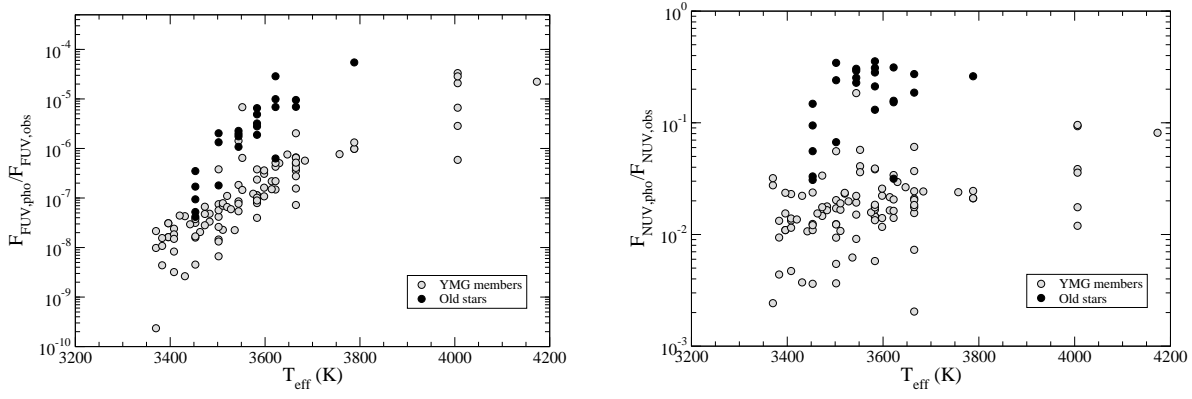}{teff_nuv_ratio_v2.eps}
			\caption{The fraction of the observed FUV (left) and NUV (right) flux density contributed by the stellar photosphere as a function of stellar effective temperature. \newline\label{teff_uvratio}}
		\end{figure}

\begin{figure}[tbp]
			\plottwo{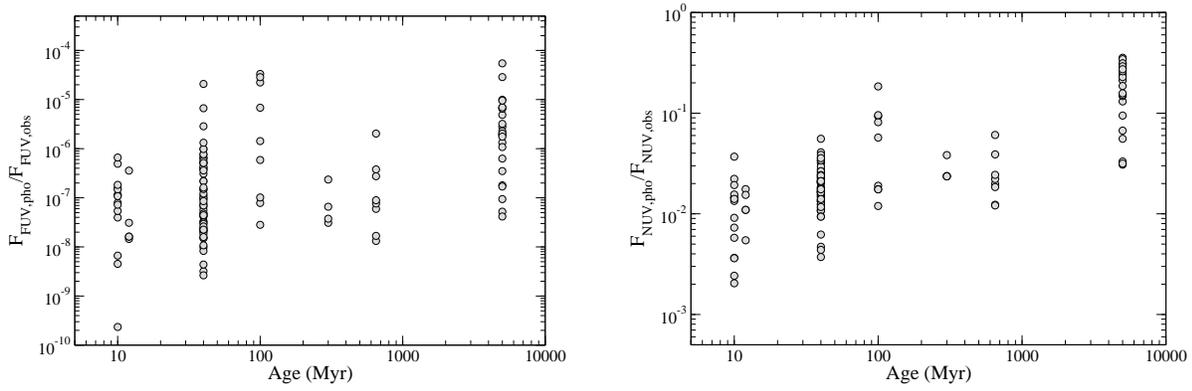}{age_nuv_ratio_v2.eps}
					\caption{The fraction of the observed FUV (left) and NUV (right) flux contributed by the stellar photosphere as a function of stellar age. \newline\newline\newline \label{age_uvratio}}
		\end{figure}

Using the stellar age and published SpT, we derive a \Teff\ and mass using the \cite{bara98} models and measure the corresponding photospheric flux from the PHOENIX models for each star in our sample. 
In the NUV, the photospheric flux comprised $<5$\% of the observed flux in most stars, except for the oldest M dwarfs in which the photospheric contribution peaks at 40\%. In the FUV, the photosphere is negligible, comprising only 0.005\% or less of the observed flux. 
The ratios of the photospheric to observed flux densities as a function of \Teff\ are shown in Figure~\ref{teff_uvratio} and as a function of age in Figure~\ref{age_uvratio}. A trend with \Teff\ in the FUV (left) plot of Figure~\ref{teff_uvratio} is evident, which is not seen in the NUV (right) plot. This difference is due to the steeper drop in photospheric FUV flux compared to the NUV with decreasing \Teff. For all stars we subtract the photospheric contribution from the observed \galex\ flux densities providing the excess emission, ($F_{FUV}/F_J$)$_{exc}$ and ($F_{NUV}/F_J$)$_{exc}$, as a measure of pure upper-atmosphere activity.

By comparing the absolute J band magnitudes of the old sample (all of which have published trigonometric distances) to the model predictions, we find a mean absolute deviation of 0.4 mag, corresponding to a 31\% uncertainty in model J-band flux density. Given that we are subtracting a relatively small value of fractional photospheric FUV and NUV flux from the observed flux, this uncertainty is not significant in the quantities reported.

\section{Results}\label{results}

\subsection{Correlations Between Stellar Activity Diagnostics}\label{correlations}

X-ray emission is ubiquitous among low-mass stars and is indicative of active stellar upper-atmospheres throughout their lifetimes, e.g.~94\% of all K and M dwarfs within 6 pc exhibited detectable X-ray emission as observed by \rosat\footnote{The R\"ontgensatellit (\rosat) was a joint German, US and British X-ray observatory operational from 1990 to 1999.} (\citealt{schm95}). Fractional X-ray luminosities have also been shown to be ``saturated'' across a wide range of spectral types, H$\alpha$ equivalent widths, and ages at the value of log($L_X/L_{\rm bol}$) $\sim -3$, with the bulk of the dispersion in both field and cluster samples between  log($L_X/L_{\rm bol}$) of --2 and --4 primarily due to variations in stellar rotation (\citealt{stau97,delf98,jack12}).

Correlations among stellar activity indicators are useful in understanding the formation mechanisms of emission features, energy distributions in the stellar atmosphere, and to allow one activity diagnostic to act as a proxy for another. Should observations of X-ray, FUV and NUV fluxes correlate with the EUV, then more accurate EUV flux estimates can be obtained.
 At the moment, there are fewer than 10 M dwarfs (with a wide range of stellar masses) for which EUV data exist in the archives making the robust connection between EUV and X-ray and between EUV and FUV/NUV impossible for a wide range of stellar masses and ages. Interpolating between X-ray and FUV fluxes is yet another possible route, assuming the line formation mechanisms are the same in all three wavelength ranges and  correlations are observed.  \cite{mitr05} observed 5 dMe stars (the classic flare stars AT Mic, AU Mic, EV Lac, UV Cet and YZ CMi) simultaneously in X-ray and UV wavelengths. They find a significant correlation from which they conclude that stellar chromospheres and coronae are both continuously heated by common impulsive energy release processes, at least for very active M dwarfs.  

In order to compare X-ray and UV data, we cross-referenced the sample of YMG M stars against the \rosat\ All-Sky Survey Bright Source Catalog and Faint Source Catalog \citep{voge99,voge00}. Our query was limited to a search radius of 38\arcsec\ around the 2MASS coordinates, the 3$\sigma$ positional error determined by \cite{voge99}. For the field sample, the search radius was increased to accommodate the very high proper motion stars.  The empirically-calibrated X-ray flux ($F_X$) in erg~s$^{-1}$~cm$^{-2}$ was calculated using the count-rate conversion equation of \cite{schm95}.

\begin{figure}[tbp]
									\plotone{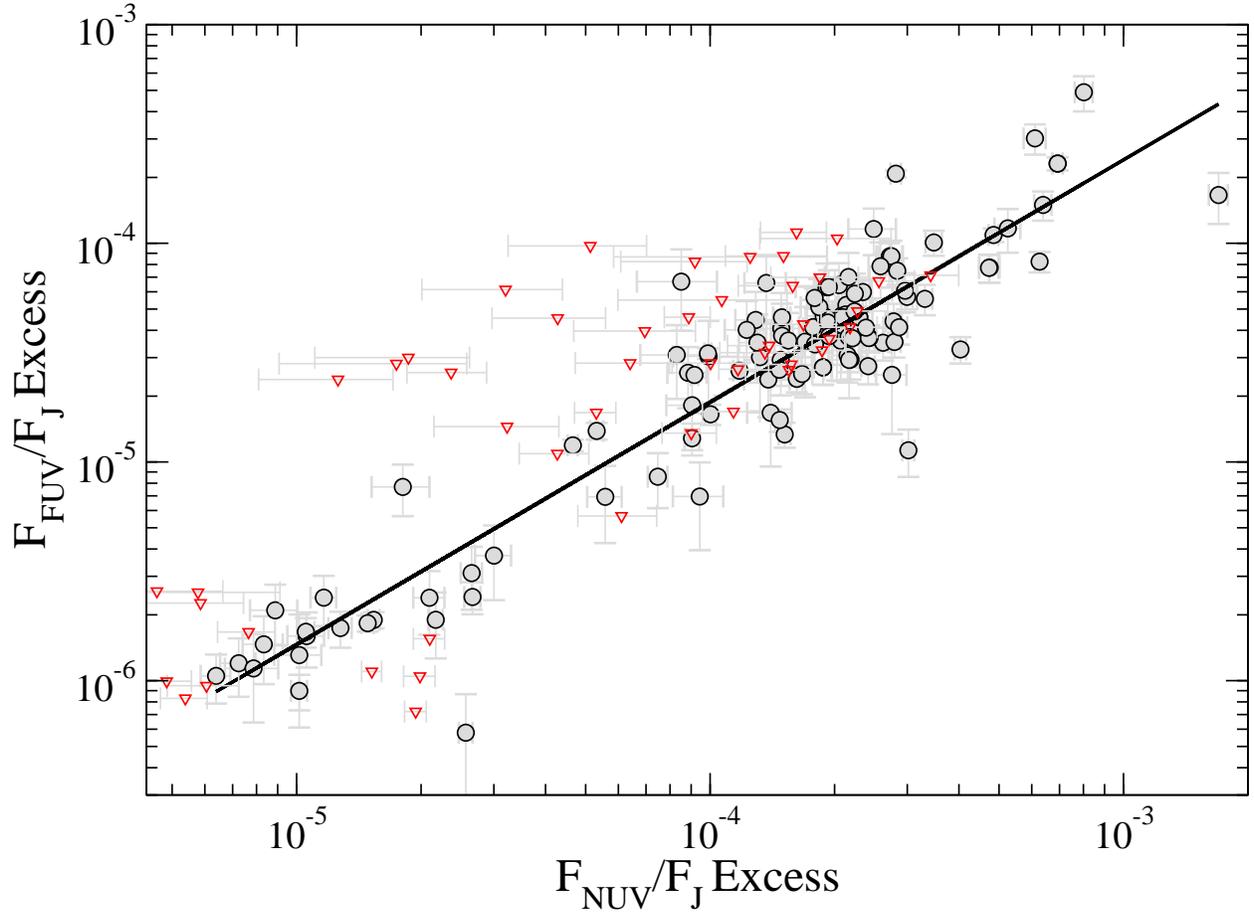}
										\caption{The \galex\ NUV and FUV excess fractional flux density for stars in our sample with detections in both bandpasses (grey dots). The solid line is a regression fit to the detections with a slope of 1.11$\pm$0.04.  The red triangles are FUV upper limits with NUV detections. \label{fuv_nuv}}
									\end{figure}

\begin{figure}[tbp]
									\plottwo{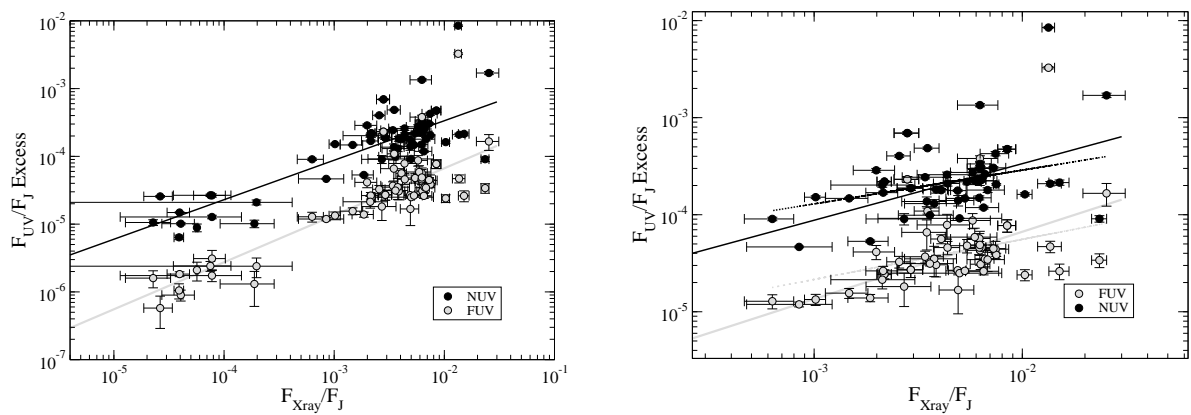}{xray_uv_exc_highend_v2.eps}
										\caption{The excess NUV and FUV fractional flux density plotted against the X-ray fractional flux from \rosat.  The coefficients of the regression lines are listed in Table~\ref{coefficients}.  The right plot focuses on the high-activity cluster of points with correlations from the left plot (solid line) plus the regression lines fit only to the shown high flux data (dashed line), showing a much weaker correlation between UV and X-ray data.  
										\label{xray_uvexc}}
									\end{figure}

 Figure~\ref{fuv_nuv} shows a correlation between $F_{FUV}/F_J$ and $F_{NUV}/F_J$ across the wide range of fluxes with a correlation coefficient $R$=0.94. Comparing the $F_{X}/F_J$ to ($F_{FUV}/F_J$)$_{exc}$ and ($F_{NUV}/F_J$)$_{exc}$, significant correlations exist using the full range of fluxes ($R=0.85$), which covers 3 orders of magnitude in X-ray and UV fluxes. As seen in Figure~\ref{xray_uvexc}, the correlation is primarily defined by two clusters: one group of low-emitters and one of high-emitters (roughly old and young stars, respectively). When focusing on only the strong and weak emitters, the correlations weaken ($R$ = 0.36 and 0.18, respectively, for the NUV and $R$ = 0.42 and 0.49 for the FUV). Results of the regression analyses are summarized in Table~\ref{coefficients}.  In almost all cases, the NUV and FUV observations were taken simultaneously by \galex, but the \rosat\ X-ray observations were collected years earlier. It is most likely that the non-simultaneity of the observations contributes to the lack of a correlation due to the short-term flaring distribution and long-term activity cycles.

\subsection{Intra-age Stellar Variability}\label{variability}

\begin{figure}[tbp]
												\includegraphics[width=5in,angle=90]{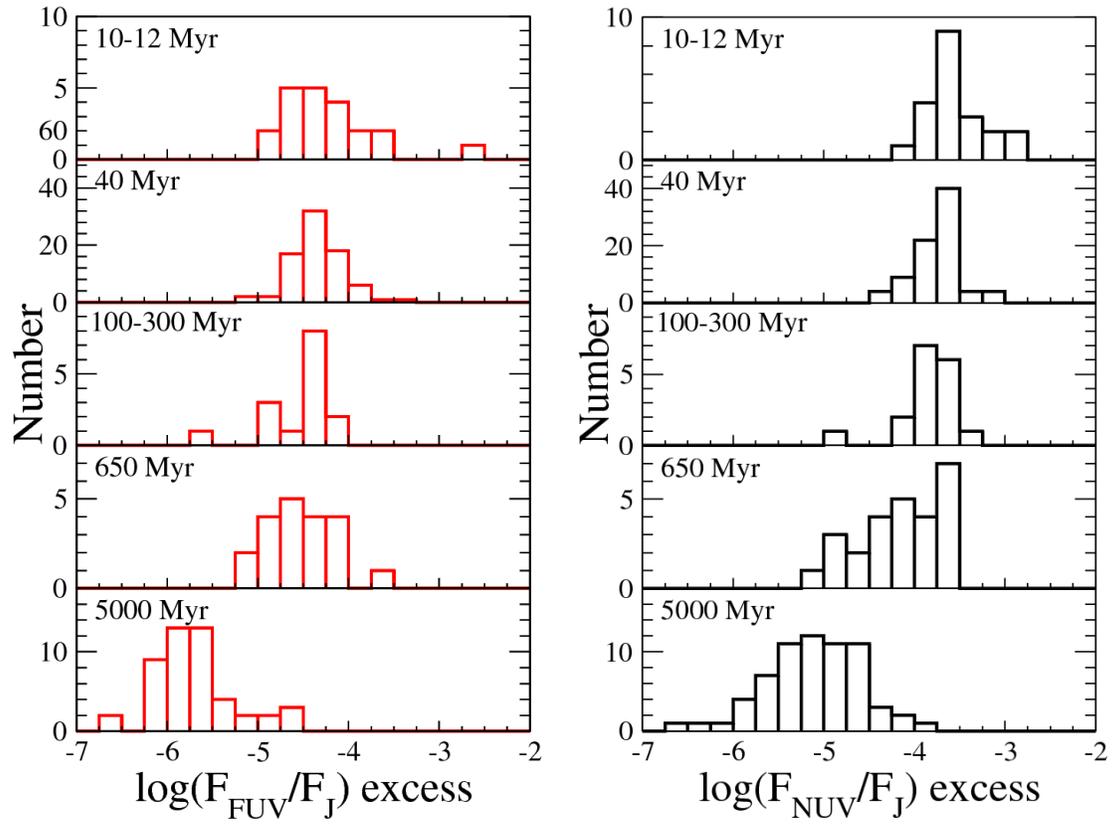}
													\caption{Histogram of excess fractional FUV (left) and NUV (right) flux densities including upper limits.
				\label{hist_uv}}
											\end{figure}

We observe a span of 1--2 orders of magnitude in UV activity at each age as seen in Figure~\ref{hist_uv}.  With the many more FUV upper limits in the Hyades and old field samples, it is likely that the full span of emission levels is even larger at these older ages. Some of this wide range in UV activity among the field stars is due to the uncertain stellar ages, from 1 to 10 Gyr. The large spread in measured rotation periods of M stars at both young and old ages (e.g. \citealt{irwi11} and reference therein) must also contribute to the UV flux variation, although \cite{pizz03} showed that young stars with saturation-level X-ray emission, do not follow the expected rotation-activity relation seen in older stars. 
Short transient events such as flares must also contribute.  \cite{wels07} observed such events in 3\% of old field M dwarfs found in the the \galex\ archive within a single 1500-s exposure, while \HST\ UV spectra of a few old M dwarf planet hosts revealed a variety of flare activity, with variability amplitudes ranging from factors of 2 to 10 on timescales of 100 -- 1000 seconds \citep{fran13}. For young Ms, larger and more frequent flares are expected.

\subsection{Activity Drop with Age}\label{activity_age}

It has been well-established that the chromospheric activity and coronal emission of FGKM stars steadily decreases with age due to the reduced dynamo production of magnetic fields as the star spins down. Unlike the spin-down time scale for higher-mass stars ($<$1 Gyr; e.g.~\citealt{skum72}), the spin-down time-scales for field M dwarfs range from 1 to 10 Gyr, taking longer with decreasing stellar mass (\citealt{delf98,irwi11}).  The angular momentum evolution of early-Ms appears to be the most dramatic between 10 and 300 Myr old, increasing rotation rates until about 100 Myr and then starting a slow decline (\citealt{irwi11}; Kidder, Shkolnik \& Skiff, in preparation).  

\begin{figure}[tbp]
											\plottwo{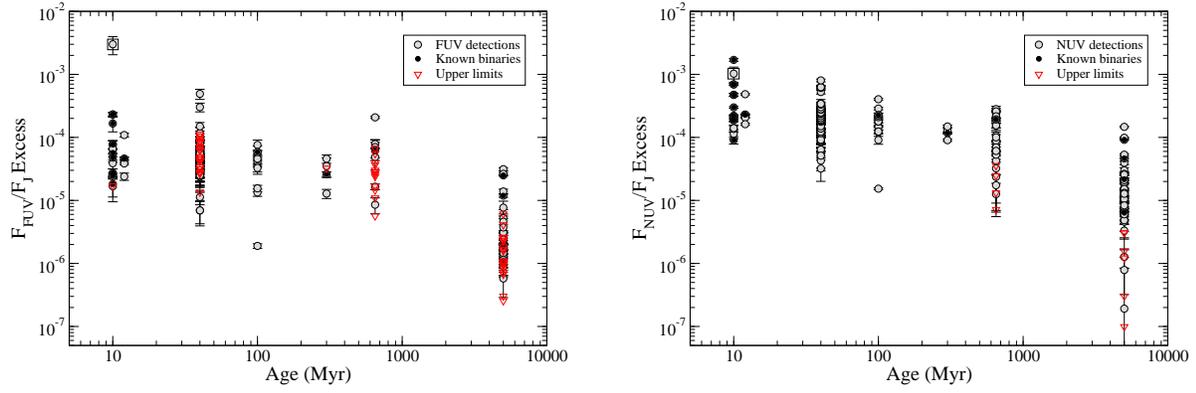}{age_nuv_v2.eps}
												\caption{The excess FUV (left) and NUV (right) fractional flux density, where the photospheric flux from the appropriate model is subtracted from the observed flux, plotted as a function of stellar age. The boxed data point is the accreting star TWA 31. \label{age_uv}}
										\end{figure}

\begin{figure}[tbp]
										\includegraphics[scale=.50,angle=270]{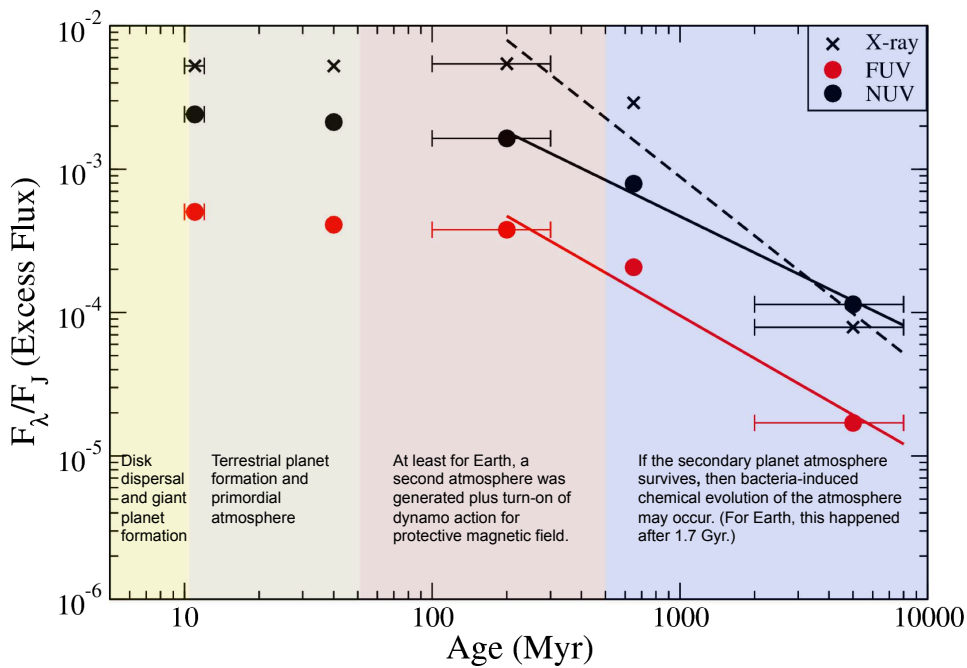}
											\caption{Median X-ray, FUV and NUV excess fluxes (not flux densities), including upper limits, as a function of stellar age. The FUV median values for the Hyades cluster and the old sample are comprised of 65\% and 42\% upper limits, respectively. Using the correlation in Figure~\ref{fuv_nuv}, we predicted the FUV values for those with upper limits. The coefficients of the power-law fits to the three bandpasses are listed in Table~\ref{coefficients}.\label{age_evolution_ppt}}
										\end{figure}

Using the relatively accurate ages of the YMG members, we map the evolution of the high FUV and NUV emission as a function of time.  Figures~\ref{age_uv} and \ref{age_evolution_ppt} shows no significant evolution in both the FUV and NUV emission from 10 to a few hundred Myrs with a decline beginning by 650 Myr followed by a sharp drop by two orders of magnitude in the old sample. 
This is similar behavior to X-ray data \citep{prei05}.  \cite{gude97} showed a $t^{-1.5}$ decline in X-ray luminosity for solar mass stars beyond 1 Gyr. A slower decrease of $L_X \propto t^{-1}$ is predicted by \cite{feig04} for lower-mass stars,  but they measured $L_X \propto t^{-2}$. With the few points they had, they could not rule out a shallower decline. 
We measure a drop in NUV and FUV fractional  flux\footnote{Note that due the high fraction (42\%) of FUV upper limits in the old sample, we use predicted values of the FUV based on the correlation found in Figure~\ref{fuv_nuv}.} to be proportional to $t^{-0.84\pm0.09}$ and $t^{-0.99\pm0.19}$, respectively, for ages $\gtrsim$200 Myr, with a decline in the fractional X-ray flux of our sample to be $t^{-1.36\pm0.32}$. (See Table~\ref{coefficients}.)
The consistency between the X-ray and the FUV implies that, at least qualitatively, 
we can draw similar conclusions for the EUV -- i.e.~a saturation level of emission until a few hundred millions of years and a reduction in flux with age afterward following roughly $t^{-1}$. The decline in the NUV is notably shallower.

Comparing the median excess fluxes of the youngest (TWA + $\beta$ Pic) to the oldest stars reveals a drop in emission by factors of 65, 30, and 20 in the X-ray, FUV and NUV, respectively,\footnote{Using the upper limits rather than predicted FUV values, the young-to-old ratio is 26, rather than 30.} implying that the decline in flux with age may steepen with shorter wavelength.  \cite{clai12} have shown a similar change with wavelength for a small sample of Sun-like stars.
Interpolating between the X-ray and NUV results to assess the decline in EUV flux is limited to between $\approx$30 to $\approx$65 due to the very large scatter in activity levels at each age of the sample.


\section{Summary}\label{summary}

Using archived \galex\ photometry, we analyzed the evolution of the FUV and NUV emission in early M stars.  Our sample consisted of  215 stars at ages of 10, 12, 40, 100, 300 and 650 Myr, probing critical planet formation and evolution time scales. These stars are confirmed members of known YMGs, the Hyades cluster and the old field sample within 10 pc. Ninety-five percent of the targets observed by \galex\ in the NUV were detected, while 184 of the stars observed in the FUV had a 68\% detection rate.  

We used current (i.e.~photosphere-only) PHOENIX models to calculate the photospheric contribution to the two \galex\ bandpasses and subtracted it from the observed quantities to study the evolution of the stellar activity originating from the chromosphere, transition region and corona. The main results from the analysis of the X-ray, FUV and NUV flux emitted from these M stars are:

\textbullet\ In most cases, the photospheric flux in the NUV bandpass contributes less than 5\% of the total observed flux, except for the oldest M dwarfs, in which the photospheric contribution peaks at $\sim$40\%. In the FUV, the photosphere contributes at most 0.005\% of the observed flux.

\textbullet\ A range of 1 to 2 orders of magnitude in UV activity is observed at each age.  Stellar rotation, unknown binarity, long-term activity cycles and short-term flaring all likely contribute to this wide range, highlighting the difficulty of extending single activity measurements of M dwarfs at UV or X-ray wavelengths to the EUV.  

\textbullet\ The X-ray and UV fluxes correlate over a broad range of activity levels, defined by two groups: high and low emitters.  Within each group there is only a weak correlation, likely due to the non-simultaneity of the UV and X-ray observations.

\textbullet\ Qualitatively, the FUV and NUV excess flux densities decay in a similar fashion to X-ray results, with a high saturation level from 10 Myr until a few hundred Myrs. By the age of the Hyades at 650 Myr, we measure a drop in excess flux,  after which it plummets at the old ages of the field sample.   Without these measurements at each individual age, a single power-law fit to just the TWA and oldest stars, as done by \cite{stel13}, underestimates the UV emission from M stars by a factor of 3--5 over many hundreds of Myrs.

\textbullet\ The median excess fractional fluxes at each age show a reduction in FUV and NUV by factors of 30 and 20, respectively, from young to old ages.  Combining this with an observed drop in X-ray by a factor of 65 suggests that the median reduction in flux with age may steepen with shorter wavelength, and that the drop in EUV flux most likely falls in between.

The reported FUV and NUV fluxes provide the empirical guidance needed to build new M dwarf upper-atmosphere models (Peacock, Barman \& Shkolnik, in preparation), to characterize of the full-UV spectra, including the EUV and \Lya, for the stars that are the most common planet hosts. These models will provide a grid of input spectra to planetary atmospheric photochemical models to study the impact of M dwarf UV evolution on planetary atmospheres.

\acknowledgements

E.S. would like to thank V.~Meadows for stimulating discussions, B.~Stelzer and the referee, L.~Fossati, for helpful comments on the manuscript, and S.~Neff and the \galex\ team for answering questions about the archive. This material is based upon work supported by the NASA/GALEX grant program under Cooperative Agreement No.~NNX12AC19G issued through the Office of Space Science, and through generous funding to Lowell Observatory by D.~Trantow and M.~Beckage. This research has made use of the VizieR catalogue access tool, CDS, Strasbourg, France \citep{ochs00} and the Mikulski Archive for Space Telescopes (MAST). STScI is operated by the Association of Universities for Research in Astronomy, Inc., under NASA contract NAS5-26555. Support for MAST for non-HST data is provided by the NASA Office of Space Science via grant NNX13AC07G and by other grants and contracts.


	\clearpage

	\clearpage

	\clearpage


		
\clearpage
\bibliography{/Users/evgenyashkolnik/Dropbox/refs_master}{}
\bibliographystyle{apj}

\clearpage

\clearpage \begin{deluxetable}{lllccccccccccc}																																																																	
\tabletypesize{\scriptsize}																																																																	
\rotate																																																																	
\tablecaption{Target Early M Stars Observed by \galex. \label{table_targets}}																																																																	
\tablewidth{0pt}																																																			
\tablehead{																										
\colhead{Name} 	&	 \colhead{R.A.$_{J2000}$} 	&	 \colhead{Dec.$_{J2000}$} 	&	 \colhead{SpT}	&	 \colhead{J$_{2MASS}$} 	&	 \colhead{Dist.\tablenotemark{a}}	&	 \colhead{Bin.\tablenotemark{b}} 	&	 \colhead{Bin.~Sep.\tablenotemark{c}} 	&		 \colhead{$F_{FUV}$}			&		 \colhead{$F_{NUV}$}			&	 \colhead{Refs.\tablenotemark{d}}	\\
\colhead{} 		&	 \colhead{(deg.)} 	&	 \colhead{(deg.)} 	&	 \colhead{} 	&	 \colhead{mag} 	&	 \colhead{pc} 	&	 \colhead{Type}	&	 \colhead{\arcsec}	&		 \colhead{$\mu$Jy} 			&		 \colhead{$\mu$Jy} 			&	 \colhead{}	
}																																																												
\startdata																												
\textbf{TW Hydra Association, 10 Myr old} \\																											
TWA 2 A	&	167.307540	&	-30.027720	&	M2	&	7.6	&	46.6	&	VB	&	0.5	&		38.18	$\pm$	5.70	&		268.61	$\pm$	10.12	&	1, 2	\\
TWA 2 B	&	167.307540	&	-30.027720	&	M2	&	7.6	&	46.6	&	VB	&	0.5	&		38.18	$\pm$	5.70	&		268.61	$\pm$	10.12	&	1, 2	\\
TWA 3 Aab	&	167.616200	&	-37.531100	&	M3	&	7.7	&	-34.0	&	VB, SB2	&	1.5	&		320.16	$\pm$	19.77	&		961.84	$\pm$	19.42	&	1, 2	\\
TWA 3 B	&	167.616200	&	-37.531100	&	M3.5	&	7.7	&	-34.0	&	VB	&	1.5	&		320.16	$\pm$	19.77	&		961.84	$\pm$	19.42	&	1, 2	\\
TWA 12	&	170.272917	&	-38.754722	&	M2	&	9.0	&	64.1	&	--	&	--	&		15.55	$\pm$	5.86	&		82.10	$\pm$	6.39	&	3, 2	\\
TWA 13 B	&	170.321830	&	-34.779310	&	M2	&	8.4	&	55.6	&	VB	&	4.9	&		52.17	$\pm$	7.12	&		321.75	$\pm$	11.91	&	4, 4	\\
TWA 13 A	&	170.322500	&	-34.780600	&	M1	&	8.4	&	55.6	&	VB	&	4.9	&		52.17	$\pm$	7.12	&		321.75	$\pm$	11.91	&	4, 4	\\
TWA 5 Aab	&	172.980250	&	-34.607567	&	M1.5	&	7.7	&	50.1	&	VB+BD, SB2	&	2.5	&		65.22	$\pm$	9.94	&		300.80	$\pm$	11.17	&	5, 6	\\
TWA 8 A	&	173.172900	&	-26.865300	&	M2-M3	&	8.3	&	-38.0	&	VB	&	13.7	&		41.84	$\pm$	4.60	&		222.24	$\pm$	4.43	&	7, 7	\\
TWA 9 B	&	177.098870	&	-37.480140	&	M1	&	10.0	&	50.3	&	VB	&	5.9	&		26.86	$\pm$	6.93	&		275.22	$\pm$	11.32	&	7, 7	\\
TWA 31\tablenotemark{d}	&	181.795375	&	-32.514922	&	M4.2	&	13.0	&	80.0	&	--	&	--	&		28.82	$\pm$	4.94	&		9.86	$\pm$	3.49	&	8	\\
TWA 23 ab	&	181.864080	&	-32.783420	&	M1	&	8.6	&	53.9	&	SB2	&	0.0	&		10.31	$\pm$	3.96	&		53.33	$\pm$	5.55	&	9, 8	\\
TWA 20 ab	&	187.908620	&	-45.983170	&	M2	&	9.3	&	77.3	&	SB2	&	0.0	&	$<$	5.00			&		34.34	$\pm$	5.20	&	10, 11	\\
TWA 16	&	188.734583	&	-45.635453	&	M1.8	&	9.0	&	78.4	&	--	&	--	&		26.42	$\pm$	7.03	&		56.14	$\pm$	6.30	&	12	\\
TWA 10	&	188.767710	&	-41.610720	&	M2.5	&	9.1	&	-53.0	&	--	&	--	&		5.98	$\pm$	3.93	&		51.00	$\pm$	5.47	&	7	\\
TWA 30 B\tablenotemark{d} 	&	173.075929	&	-30.308792	&	M4	&	15.4	&	-42.0	&	VB	&	80.0	&	$<$	5.00			&	$<$	2.50			&	13, 13	\\
\hline																																																									
\textbf{$\beta$ Pic YMG, 12 Myr old}\\																											
HIP 23418	&	75.494991	&	+09.983098	&	M3	&	7.2	&	32.1	&	--	&	--	&		226.04	$\pm$	14.56	&		1010.39	$\pm$	16.34	&	3	\\
AT Mic B	&	310.462920	&	-32.436110	&	M4	&	5.8	&	10.2	&	VB	&	3.2	&		347.01	$\pm$	15.70	&		1754.65	$\pm$	14.87	&	3	\\
AT Mic A	&	310.463350	&	-32.435280	&	M4	&	5.8	&	10.2	&	VB	&	3.2	&		347.01	$\pm$	15.70	&		1754.65	$\pm$	14.87	&	3	\\
AU Mic	&	311.289580	&	-31.340830	&	M1	&	5.4	&	9.9	&	--	&	--	&		410.54	$\pm$	13.47	&		2221.25	$\pm$	19.69	&	3	\\
LP 984-91	&	341.241509	&	-33.250484	&	M4.0	&	7.8	&	23.6	&	VB	&	35.5	&		29.28	$\pm$	3.55	&		201.09	$\pm$	5.92	&	9 \& 14, 14	\\
\hline																																																									
\textbf{Tuc-Hor YMG. 40 Myr old}\\																											
HIP 1910	&	6.037500	&	-62.184440	&	M1	&	8.4	&	46.3	&	--	&	--	&		33.07	$\pm$	5.69	&		152.52	$\pm$	7.50	&	3	\\
CT Tuc	&	6.311250	&	-61.513330	&	M1	&	8.6	&	37.5	&	--	&	--	&		16.72	$\pm$	2.68	&		127.04	$\pm$	4.92	&	3	\\
HIP 3556	&	11.367317	&	-51.626090	&	M3	&	8.5	&	38.5	&	--	&	--	&		4.46	$\pm$	2.04	&		38.03	$\pm$	3.23	&	3	\\
GSC 8056-0482	&	39.215438	&	-52.051011	&	M3	&	8.4	&	-25.0	&	--	&	--	&		23.91	$\pm$	3.95	&		180.77	$\pm$	5.70	&	3	\\
AF Hor	&	40.447080	&	-52.991940	&	M2	&	8.5	&	-42.0	&	--	&	--	&		30.20	$\pm$	3.59	&		136.69	$\pm$	5.31	&	3	\\
HIP 107345	&	326.125430	&	-60.977500	&	M1	&	8.8	&	42.3	&	--	&	--	&		18.53	$\pm$	2.41	&		124.33	$\pm$	3.18	&	3	\\
TYC 9344-0293	&	351.544590	&	-73.397220	&	M0	&	8.8	&	-46.0	&	--	&	--	&		40.37	$\pm$	6.26	&		129.09	$\pm$	7.36	&	3	\\
04133314-5231586	&	63.388090	&	-52.532970	&	M1.7	&	10.0	&	-47.2	&	--	&	--	&		Not. Obs.			&		4.26	$\pm$	1.87	&	15	\\
00514081-5913320	&	12.920050	&	-59.225580	&	M4.1	&	11.3	&	-41.3	&	--	&	--	&	$<$	3.00			&		1.70	$\pm$	0.93	&	15	\\
03114544-4719501	&	47.939350	&	-47.330590	&	M3.7	&	10.4	&	-33.1	&	--	&	--	&	$<$	3.00			&		7.11	$\pm$	1.90	&	15	\\
01024375-6235344	&	15.682300	&	-62.592910	&	M3.8	&	9.6	&	-22.0	&	--	&	--	&	$<$	3.00			&		20.60	$\pm$	2.51	&	15	\\
23170011-7432095	&	349.250460	&	-74.535980	&	M4.1	&	10.4	&	-27.8	&	--	&	--	&	$<$	5.00			&		5.01	$\pm$	2.61	&	15	\\
04470041-5134405	&	71.751720	&	-51.577920	&	M2.4	&	10.1	&	-41.9	&	--	&	--	&		4.64	$\pm$	1.67	&		13.02	$\pm$	2.43	&	15	\\
00485254-6526330	&	12.218950	&	-65.442520	&	M3.2	&	10.4	&	-39.1	&	--	&	--	&	$<$	5.00			&		10.01	$\pm$	1.34	&	15	\\
21504048-5113380	&	327.668680	&	-51.227250	&	M4.2	&	10.3	&	-25.9	&	--	&	--	&		3.53	$\pm$	2.09	&		11.84	$\pm$	2.07	&	15	\\
21275054-6841033	&	321.960620	&	-68.684250	&	M4.1	&	10.4	&	-28.1	&	--	&	--	&		Not. Obs.			&		11.99	$\pm$	0.84	&	15	\\
00332438-5116433	&	8.351620	&	-51.278710	&	M2.4	&	9.9	&	-38.5	&	--	&	--	&		1.26	$\pm$	1.48	&		17.71	$\pm$	2.50	&	15	\\
02341866-5128462	&	38.577790	&	-51.479510	&	M4.2	&	10.6	&	-29.4	&	--	&	--	&		5.94	$\pm$	1.55	&		7.83	$\pm$	1.96	&	15	\\
02045317-5346162	&	31.221570	&	-53.771180	&	M3.4	&	10.4	&	-36.5	&	--	&	--	&	$<$	3.00			&		10.94	$\pm$	2.19	&	15	\\
23570417-0337559	&	359.267450	&	-03.632330	&	M3.3	&	10.9	&	-46.8	&	--	&	--	&		1.77	$\pm$	0.53	&		6.35	$\pm$	0.77	&	15	\\
01283025-4921094	&	22.126080	&	-49.352630	&	M4.0	&	10.6	&	-31.3	&	--	&	--	&	$<$	3.00			&		13.15	$\pm$	2.25	&	15	\\
02321934-5746117	&	38.080610	&	-57.769940	&	M3.8	&	11.1	&	-42.7	&	--	&	--	&	$<$	5.00			&		7.39	$\pm$	3.45	&	15	\\
01160045-6747311	&	19.001910	&	-67.791980	&	M4.2	&	11.8	&	-49.4	&	--	&	--	&	$<$	3.00			&		1.67	$\pm$	1.28	&	15	\\
02590284-6120000	&	44.761850	&	-61.333350	&	M3.5	&	11.6	&	-60.3	&	--	&	--	&	$<$	3.00			&		3.45	$\pm$	1.50	&	15	\\
00394063-6224125	&	9.919300	&	-62.403480	&	M4.2	&	11.2	&	-39.1	&	--	&	--	&		Not. Obs.			&		7.57	$\pm$	1.78	&	15	\\
02001992-6614017	&	30.083040	&	-66.233830	&	M3.0	&	10.7	&	-49.4	&	--	&	--	&		3.28	$\pm$	1.99	&		12.20	$\pm$	1.38	&	15	\\
01275875-6032243	&	21.994810	&	-60.540090	&	M4.0	&	11.1	&	-39.6	&	--	&	--	&	$<$	2.00			&		8.33	$\pm$	1.63	&	15	\\
04213904-7233562	&	65.412690	&	-72.565610	&	M2.4	&	9.9	&	-37.8	&	--	&	--	&	$<$	5.00			&		28.80	$\pm$	2.93	&	15	\\
02543316-5108313	&	43.638180	&	-51.142050	&	M1.4	&	8.7	&	-27.3	&	--	&	--	&		20.44	$\pm$	3.32	&		83.31	$\pm$	4.52	&	15	\\
00273330-6157169	&	6.888760	&	-61.954720	&	M3.5	&	10.3	&	-33.9	&	--	&	--	&	$<$	5.00			&		20.07	$\pm$	3.38	&	15	\\
00152752-6414545	&	3.864670	&	-64.248500	&	M1.5	&	9.3	&	-36.3	&	--	&	--	&		13.26	$\pm$	3.27	&		39.69	$\pm$	4.51	&	15	\\
22223966-6303258	&	335.665290	&	-63.057180	&	M3.2	&	10.2	&	-35.3	&	--	&	--	&		4.77	$\pm$	2.25	&		23.28	$\pm$	3.81	&	15	\\
02242453-7033211	&	36.102240	&	-70.555890	&	M4.0	&	10.4	&	-28.3	&	--	&	--	&	$<$	3.00			&		13.53	$\pm$	1.65	&	15	\\
03291649-3702502	&	52.318740	&	-37.047280	&	M3.7	&	10.7	&	-36.1	&	--	&	--	&		2.63	$\pm$	1.59	&		11.79	$\pm$	3.10	&	15	\\
00493566-6347416	&	12.398610	&	-63.794900	&	M1.8	&	9.3	&	-33.6	&	--	&	--	&		9.04	$\pm$	2.76	&		46.93	$\pm$	3.59	&	15	\\
23291752-6749598	&	352.323020	&	-67.833300	&	M3.9	&	10.8	&	-35.6	&	--	&	--	&		3.08	$\pm$	1.46	&		9.64	$\pm$	1.36	&	15	\\
02502222-6545552	&	42.592620	&	-65.765360	&	M2.8	&	10.3	&	-42.1	&	--	&	--	&		2.87	$\pm$	2.07	&		17.07	$\pm$	2.36	&	15	\\
02070176-4406380	&	31.757340	&	-44.110570	&	M1.2	&	9.3	&	-37.5	&	--	&	--	&		7.86	$\pm$	1.97	&		53.39	$\pm$	2.84	&	15	\\
04133609-4413325	&	63.400410	&	-44.225700	&	M3.3	&	10.77	&	-44.5	&	--	&	--	&	$<$	5.00			&		12.63	$\pm$	3.21	&	15	\\
21143354-4213528	&	318.639790	&	-42.231350	&	M4.1	&	11.4	&	-43.5	&	--	&	--	&	$<$	5.00			&		7.34	$\pm$	1.27	&	15	\\
00235732-5531435	&	5.988860	&	-55.528770	&	M4.0	&	11.1	&	-40.4	&	--	&	--	&	$<$	5.00			&		8.77	$\pm$	2.61	&	15	\\
02474639-5804272	&	41.943320	&	-58.074220	&	M1.6	&	9.4	&	-35.5	&	--	&	--	&		11.85	$\pm$	2.25	&		51.98	$\pm$	3.65	&	15	\\
00421010-5444431	&	10.542090	&	-54.745310	&	M3.0	&	9.8	&	-31.6	&	--	&	--	&		8.80	$\pm$	2.38	&		36.27	$\pm$	2.89	&	15	\\
22244102-7724036	&	336.170920	&	-77.401020	&	M4.0	&	11.4	&	-45.9	&	--	&	--	&	$<$	3.00			&		8.04	$\pm$	1.79	&	15	\\
02001277-0840516	&	30.053270	&	-08.681140	&	M2.0	&	8.8	&	-25.4	&	--	&	--	&		22.34	$\pm$	0.96	&		96.91	$\pm$	1.68	&	15	\\
01224511-6318446	&	20.687960	&	-63.312390	&	M3.3	&	9.8	&	-28.7	&	--	&	--	&		6.65	$\pm$	3.05	&		39.08	$\pm$	3.48	&	15	\\
22463471-7353504	&	341.644660	&	-73.897350	&	M3.2	&	9.66	&	-27.5	&	--	&	--	&		Not. Obs.			&		42.93	$\pm$	1.16	&	15	\\
22021626-4210329	&	330.567750	&	-42.175830	&	M0.8	&	8.93	&	-33.3	&	--	&	--	&		18.48	$\pm$	2.59	&		84.09	$\pm$	2.97	&	15	\\
01351393-0712517	&	23.808040	&	-07.214380	&	M4.1	&	8.96	&	-14.2	&	--	&	--	&		17.44	$\pm$	3.60	&		87.83	$\pm$	4.65	&	15	\\
01505688-5844032	&	27.737030	&	-58.734230	&	M2.9	&	9.54	&	-28.3	&	--	&	--	&		9.13	$\pm$	2.74	&		47.93	$\pm$	3.76	&	15	\\
02205139-5823411	&	35.214130	&	-58.394760	&	M3.2	&	9.7	&	-27.8	&	--	&	--	&		10.99	$\pm$	3.64	&		40.23	$\pm$	3.95	&	15	\\
02505959-3409050	&	42.748330	&	-34.151410	&	M3.8	&	10.5	&	-32.5	&	--	&	--	&	$<$	5.00			&		23.52	$\pm$	3.54	&	15	\\
00125703-7952073	&	3.237660	&	-79.868700	&	M3.4	&	9.7	&	-25.4	&	--	&	--	&		Not. Obs.			&		46.29	$\pm$	4.08	&	15	\\
23474694-6517249	&	356.945610	&	-65.290260	&	M1.5	&	9.1	&	-32.2	&	--	&	--	&		13.60	$\pm$	3.97	&		80.38	$\pm$	5.55	&	15	\\
03050976-3725058	&	46.290700	&	-37.418280	&	M2.5	&	9.54	&	-31.6	&	--	&	--	&		7.45	$\pm$	2.30	&		53.02	$\pm$	3.85	&	15	\\
03210395-6816475	&	50.266470	&	-68.279870	&	M3.4	&	10.36	&	-33.4	&	--	&	--	&		7.18	$\pm$	2.51	&		22.15	$\pm$	3.52	&	15	\\
22025453-6440441	&	330.727230	&	-64.678920	&	M2.1	&	9.1	&	-28.3	&	--	&	--	&		13.90	$\pm$	2.84	&		84.82	$\pm$	4.16	&	15	\\
21163528-6005124	&	319.147040	&	-60.086780	&	M3.9	&	10.2	&	-27.0	&	--	&	--	&		6.97	$\pm$	2.21	&		28.90	$\pm$	2.52	&	15	\\
20423672-5425263	&	310.653010	&	-54.423990	&	M3.9	&	10.8	&	-35.2	&	--	&	--	&		5.13	$\pm$	2.00	&		16.56	$\pm$	1.86	&	15	\\
00240899-6211042	&	6.037490	&	-62.184520	&	M0.2	&	8.4	&	-28.6	&	--	&	--	&		33.07	$\pm$	5.69	&		152.52	$\pm$	7.50	&	15	\\
23452225-7126505	&	356.342720	&	-71.447380	&	M3.8	&	10.2	&	-28.3	&	--	&	--	&		3.91	$\pm$	1.77	&		29.35	$\pm$	3.05	&	15	\\
01211297-6117281	&	20.304080	&	-61.291150	&	M4.1	&	11.3	&	-42.3	&	--	&	--	&	$<$	5.00			&		9.78	$\pm$	3.52	&	15	\\
23273447-8512364	&	351.893630	&	-85.210130	&	M4.0	&	10.9	&	-35.8	&	--	&	--	&	$<$	3.00			&		16.04	$\pm$	2.52	&	15	\\
01521830-5950168	&	28.076260	&	-59.838000	&	M1.6	&	8.9	&	-30.3	&	--	&	--	&		25.25	$\pm$	3.68	&		100.47	$\pm$	4.54	&	15	\\
21443012-6058389	&	326.125500	&	-60.977500	&	M0.0	&	8.8	&	-34.8	&	--	&	--	&		18.53	$\pm$	2.41	&		124.33	$\pm$	3.18	&	15	\\
03315564-4359135	&	52.981850	&	-43.987110	&	M0.0	&	8.3	&	-28.4	&	--	&	--	&		20.84	$\pm$	4.02	&		187.51	$\pm$	6.19	&	15	\\
00144767-6003477	&	3.698650	&	-60.063260	&	M3.5	&	9.71	&	-25.1	&	--	&	--	&		8.53	$\pm$	3.42	&		50.09	$\pm$	4.72	&	15	\\
21370885-6036054	&	324.286880	&	-60.601520	&	M3.6	&	9.6	&	-23.4	&	--	&	--	&		9.72	$\pm$	2.56	&		62.20	$\pm$	2.36	&	15	\\
02125819-5851182	&	33.242490	&	-58.855060	&	M3.5	&	9.3	&	-21.0	&	--	&	--	&		24.21	$\pm$	5.36	&		81.45	$\pm$	4.96	&	15	\\
23261069-7323498	&	351.544540	&	-73.397190	&	M1.5	&	8.8	&	-28.7	&	--	&	--	&		40.37	$\pm$	6.26	&		129.09	$\pm$	7.36	&	15	\\
02420404-5359000	&	40.516860	&	-53.983350	&	M3.9	&	10.1	&	-26.7	&	--	&	--	&		16.68	$\pm$	4.24	&		36.13	$\pm$	4.68	&	15	\\
05392505-4245211	&	84.854380	&	-42.755890	&	M1.8	&	9.5	&	-36.1	&	--	&	--	&		6.60	$\pm$	3.28	&		73.66	$\pm$	5.67	&	15	\\
23285763-6802338	&	352.240160	&	-68.042750	&	M2.9	&	9.3	&	-25.4	&	--	&	--	&		23.52	$\pm$	3.25	&		90.08	$\pm$	2.62	&	15	\\
23131671-4933154	&	348.319620	&	-49.554300	&	M4.1	&	9.76	&	-20.7	&	--	&	--	&		12.03	$\pm$	2.56	&		59.29	$\pm$	3.52	&	15	\\
02423301-5739367	&	40.637550	&	-57.660200	&	K7.1	&	8.6	&	-40.7	&	--	&	--	&		21.18	$\pm$	3.40	&		173.90	$\pm$	5.23	&	15	\\
02303239-4342232	&	37.635000	&	-43.706470	&	K7.0	&	8.02	&	-32.4	&	--	&	--	&		11.13	$\pm$	3.53	&		308.04	$\pm$	8.94	&	15	\\
00393579-3816584	&	9.899160	&	-38.282890	&	M1.8	&	8.8	&	-26.3	&	--	&	--	&		27.23	$\pm$	3.91	&		163.64	$\pm$	6.08	&	15	\\
04365738-1613065	&	69.239100	&	-16.218490	&	M3.0	&	9.1	&	-22.9	&	--	&	--	&		36.06	$\pm$	4.53	&		125.45	$\pm$	5.10	&	15	\\
21100614-5811483	&	317.525620	&	-58.196750	&	M3.8	&	10.9	&	-39.6	&	--	&	--	&	$<$	5.00			&		24.12	$\pm$	3.89	&	15	\\
04074372-6825111	&	61.932190	&	-68.419770	&	M2.6	&	10.4	&	-46.6	&	--	&	--	&		12.74	$\pm$	2.71	&		57.57	$\pm$	3.89	&	15	\\
23382851-6749025	&	354.618810	&	-67.817370	&	M3.9	&	10.9	&	-37.8	&	--	&	--	&		20.75	$\pm$	3.12	&		42.17	$\pm$	2.27	&	15	\\
03244056-3904227	&	51.169030	&	-39.072980	&	M4.1	&	9.9	&	-21.7	&	--	&	--	&		26.80	$\pm$	3.85	&		114.95	$\pm$	4.85	&	15	\\
04000382-2902165	&	60.015920	&	-29.037930	&	K7.2	&	8.0	&	-30.3	&	--	&	--	&		85.70	$\pm$	8.31	&		663.68	$\pm$	14.71	&	15	\\
05111098-4903597	&	77.795750	&	-49.066600	&	M3.7	&	10.6	&	-35.8	&	--	&	--	&		43.23	$\pm$	7.13	&		70.95	$\pm$	3.20	&	15	\\
\hline																																																									
\textbf{AB Dor YMG. 100 Myr old}\\																											
CD-61 1439	&	99.958340	&	-61.478330	&	K7	&	7.3	&	21.9	&	--	&	--	&		25.51	$\pm$	3.15	&		319.08	$\pm$	6.79	&	3	\\
BD+01 2447	&	157.231670	&	+00.841110	&	M2.5	&	6.2	&	7.2	&	--	&	--	&		10.20	$\pm$	0.81	&		101.36	$\pm$	1.73	&	16	\\
HD 201919	&	318.272090	&	-17.486940	&	K6	&	8.3	&	22.6	&	--	&	--	&		30.07	$\pm$	4.76	&		227.13	$\pm$	6.63	&	6	\\
HD 217379	&	345.116670	&	-26.311940	&	K7	&	7.0	&	30.0	&	--	&	--	&		37.48	$\pm$	4.16	&		392.79	$\pm$	5.32	&	3	\\
G 132-51 B (W)  	&	15.925539	&	+40.854309	&	M2.6	&	--	&	31.3	&	VB	&	0.6	&		12.97	$\pm$	3.59	&		45.96	$\pm$	5.02	&	14, 14	\\
G 132-51 B (E) 	&	15.925539	&	+40.854309	&	M3.8	&	--	&	31.3	&	VB	&	0.6	&		12.97	$\pm$	3.59	&		45.96	$\pm$	5.02	&	14, 14	\\
GJ 3136                 	&	32.223427	&	+49.449021	&	M2.9	&	8.4	&	15.5	&	--	&	--	&		18.12	$\pm$	5.74	&		124.20	$\pm$	5.92	&	14, 14	\\
G 80-21                  	&	56.847252	&	-01.972217	&	M2.8	&	7.8	&	16.3	&	--	&	--	&		Not. Obs.			&		150.45	$\pm$	8.61	&	17 \& 14	\\
NLTT 14116               	&	73.101689	&	-16.822761	&	M3.3	&	7.7	&	16.3	&	--	&	--	&		82.56	$\pm$	7.49	&		232.17	$\pm$	7.31	&	14	\\
NLTT 15049 A (SW)	&	81.423609	&	-09.153466	&	M3.8	&	8.5	&	20.7	&	VB	&	0.5	&		Not. Obs.			&		61.96	$\pm$	7.89	&	14, 14 \& 18	\\
BD+20 1790	&	110.931622	&	+20.416288	&	K7	&	7.6	&	25.8	&	--	&	--	&		46.40	$\pm$	6.22	&		568.68	$\pm$	12.67	&	14	\\
GJ 4231                    	&	328.043385	&	+05.626629	&	M2.4	&	8.2	&	31.8	&	--	&	--	&		32.29	$\pm$	5.68	&		177.54	$\pm$	7.94	&	14	\\
GJ 9809                  	&	346.520188	&	+63.926213	&	M0.3	&	7.8	&	24.9	&	--	&	--	&		Not. Obs.			&		276.41	$\pm$	13.96	&	19 \& 14	\\
1RXS J235133.3+312720 AB	&	357.890302	&	+31.456391	&	M2.0	&	9.8	&	45.0	&	VB (with BD)	&	2.4	&		11.71	$\pm$	2.59	&		42.71	$\pm$	3.15	&	14, 20	\\
HIP 110526	&	56.847252	&	-01.972217	&	M3	&	7.8	&	16.3	&	--	&	--	&		Not. Obs.			&		153.15	$\pm$	8.13	&	3	\\
\hline																																																									
\textbf{Ursa Major Association, 300 Myr old}\\																											
2MASS J10364483+1521394 A (N)    	&	159.186857	&	+15.360932	&	M4	&	8.7	&	20.1	&	VB	&	1.0	&		14.51	$\pm$	1.07	&		60.77	$\pm$	1.53	&	14, 14	\\
HD 95650	&	165.659764	&	+21.967136	&	M2	&	6.5	&	11.7	&	--	&	--	&		53.21	$\pm$	8.27	&		367.94	$\pm$	11.46	&	21	\\
1RXS J111300.1+102518	&	168.252503	&	+10.418295	&	M3.0	&	10.0	&	23.0	&	--	&	--	&	$<$	5.00			&		29.33	$\pm$	4.09	&	14	\\
G 10-52                    	&	177.147885	&	+07.694533	&	M3.5	&	9.5	&	20.7	&	--	&	--	&		11.33	$\pm$	3.11	&		34.35	$\pm$	3.17	&	14	\\
GJ 4381 AB                 	&	359.457703	&	+38.629529	&	M2.8	&	8.7	&	-21.3	&	VB	&	0.5	&		31.62	$\pm$	3.74	&		81.09	$\pm$	3.28	&	14, 14	\\
\hline																																																									
\textbf{Hyades Cluster, 650 Myr old}\\																											
LP 247-13	&	48.907641	&	+37.403984	&	M2.7	&	9.3	&	34.4	&	--	&	--	&		15.15	$\pm$	5.55	&		67.99	$\pm$	6.08	&	14	\\
1RXS J032230.7+285852	&	50.631912	&	+28.974766	&	M4.0	&	10.8	&	46.7	&	--	&	--	&	$<$	5.00			&		19.28	$\pm$	4.32	&	14	\\
Cl* Melotte 25 REID 187  	&	65.849178	&	+14.427931	&	M2.5	&	10.5	&	46.3	&	--	&	--	&		1.86	$\pm$	1.24	&		16.27	$\pm$	1.04	&	22	\\
Cl* Melotte 25 HAN 366   	&	66.460166	&	+15.002594	&	M4 	&	11.6	&	46.3	&	--	&	--	&		Not. Obs.			&		2.49	$\pm$	0.73	&	22	\\
TYC 1265-1118-1          	&	66.519614	&	+15.041349	&	M1	&	9.3	&	46.3	&	--	&	--	&		1.15	$\pm$	0.75	&		24.75	$\pm$	1.05	&	22	\\
Cl* Melotte 25 REID 228  	&	66.517985	&	+17.120720	&	M3 	&	10.9	&	46.3	&	--	&	--	&		15.23	$\pm$	0.45	&		20.00	$\pm$	0.36	&	22	\\
V1102 Tau	&	67.119925	&	+17.695940	&	M1             	&	8.6	&	46.3	&	VB	&	1.6	&		35.85	$\pm$	7.00	&		115.38	$\pm$	8.34	&	22, 23	\\
Cl* Melotte 25 REID 277  	&	67.477499	&	+16.912815	&	M2 	&	--	&	46.3	&	--	&	--	&	$<$	1.00			&		1.63	$\pm$	0.38	&	22	\\
V484 Tau    	&	67.599799	&	+17.499760	&	M3.5           	&	10.4	&	46.3	&	--	&	--	&	$<$	3.00			&		18.03	$\pm$	2.72	&	22	\\
GSC 01269-00867          	&	67.871049	&	+17.718683	&	M1 	&	9.1	&	46.3	&	--	&	--	&		Not. Obs.			&		35.25	$\pm$	4.67	&	22	\\
Cl* Melotte 25 REID 306  	&	68.033131	&	+17.664526	&	M2             	&	10.9	&	46.3	&	--	&	--	&		Not. Obs.			&	$<$	2.00			&	22	\\
Cl* Melotte 25 HAN 530   	&	68.120746	&	+17.904619	&	M4 	&	11.2	&	46.3	&	--	&	--	&		Not. Obs.			&		16.71	$\pm$	3.53	&	22	\\
Cl* Melotte 25 HAN 172   	&	64.448646	&	+13.661744	&	M1             	&	9.4	&	46.3	&	VB	&	0.8	&	$<$	3.00			&		13.08	$\pm$	2.10	&	22, 23	\\
Melotte 25 HAN 192  	&	64.695927	&	+13.366275	&	M1	&	9.1	&	46.3	&	--	&	--	&	$<$	6.00			&		20.69	$\pm$	2.50	&	22	\\
Melotte 25 REID 189	&	65.928942	&	+15.880928	&	M2             	&	10.7	&	46.3	&	--	&	--	&	$<$	2.00			&		1.40	$\pm$	0.75	&	22	\\
LP 415-1582              	&	69.017367	&	+18.888599	&	M3.5           	&	9.8	&	46.3	&	--	&	--	&		26.31	$\pm$	4.39	&		51.27	$\pm$	2.89	&	22	\\
LP 415-1619              	&	69.162224	&	+18.615774	&	M3 	&	9.8	&	46.3	&	--	&	--	&	$<$	5.00			&		5.30	$\pm$	1.58	&	22	\\
LP 15-292                	&	69.727959	&	+19.182240	&	M3 	&	10.2	&	46.3	&	--	&	--	&	$<$	5.00			&		27.03	$\pm$	2.08	&	22	\\
Cl Melotte 25 310        	&	70.052960	&	+19.286110	&	M1 	&	9.9	&	46.3	&	--	&	--	&	$<$	5.00			&		3.91	$\pm$	1.20	&	22	\\
Cl* Melotte 25 REID 122  	&	64.005982	&	+16.983133	&	M1 	&	10.8	&	46.3	&	--	&	--	&	$<$	3.00			&		5.62	$\pm$	2.12	&	22	\\
Cl* Melotte 25 REID 132  	&	64.180400	&	+16.822313	&	M3 	&	11.2	&	46.3	&	--	&	--	&	$<$	3.00			&		6.01	$\pm$	2.72	&	22	\\
Cl* Melotte 25 REID 142  	&	64.479131	&	+16.544498	&	M2             	&	10.0	&	46.3	&	--	&	--	&		7.04	$\pm$	3.12	&		34.11	$\pm$	3.47	&	22	\\
GJ 3290                  	&	66.819316	&	+17.241827	&	M1.5           	&	9.7	&	46.3	&	--	&	--	&	$<$	3.00			&		7.68	$\pm$	3.08	&	22	\\
Cl* Melotte 25 VA 559	&	67.482176	&	+16.914061	&	M2 	&	9.5	&	46.3	&	--	&	--	&		3.45	$\pm$	0.42	&		25.69	$\pm$	0.40	&	22	\\
LP 358-534               	&	65.482731	&	+23.418280	&	M4 	&	9.9	&	46.3	&	--	&	--	&	$<$	1.00			&		11.30	$\pm$	1.78	&	22	\\
LP 358-724               	&	66.325686	&	+23.060841	&	M4 	&	10.3	&	46.3	&	--	&	--	&		Not. Obs.			&	$<$	2.00			&	22	\\
Cl* Melotte 25 REID 135  	&	64.226897	&	+16.356930	&	M4 	&	10.4	&	46.3	&	--	&	--	&	$<$	3.00			&		2.16	$\pm$	2.01	&	22	\\
1RXS J041755.6+163249    	&	64.479131	&	+16.544498	&	M3 	&	10.0	&	46.3	&	--	&	--	&		14.08	$\pm$	4.25	&		31.42	$\pm$	2.76	&	22	\\
Cl* Melotte 25 REID 176  	&	65.664747	&	+18.269377	&	M0.5           	&	9.6	&	46.3	&	--	&	--	&		Not. Obs.			&		2.15	$\pm$	1.79	&	22	\\
1RXS J042829.4+174138    	&	67.122875	&	+17.694851	&	M2 	&	--	&	46.3	&	--	&	--	&		35.85	$\pm$	7.00	&		115.38	$\pm$	8.34	&	22	\\
\hline																																																									
\textbf{Field sample, $\sim$5 Gyr old}\\																											
Gl  1       	&	1.351785	&	-37.357361	&	 M1.5	&	5.3	&	4.3	&	--	&	--	&	$<$	3.00			&		49.93	$\pm$	2.38	&	24	\\
Gl  48      	&	15.634300	&	+71.679816	&	 M3.0	&	6.3	&	8.2	&	--	&	--	&		6.62	$\pm$	2.16	&		66.77	$\pm$	4.64	&	24	\\
Gl  49      	&	15.661952	&	+62.345048	&	 M1.5	&	6.2	&	10.0	&	--	&	--	&		Not. Obs.			&		226.49	$\pm$	11.82	&	24	\\
Gl  54      	&	17.595428	&	-67.444959	&	 M2.0	&	6.0	&	8.2	&	--	&	--	&		7.42	$\pm$	2.27	&		71.25	$\pm$	3.97	&	24	\\
Gl  84      	&	31.270188	&	-17.614637	&	 M2.5	&	6.5	&	9.1	&	--	&	--	&		10.47	$\pm$	2.68	&		52.20	$\pm$	5.06	&	24	\\
GJ 3135     	&	31.452312	&	-30.176639	&	 M2.5	&	8.4	&	9.3	&	--	&	--	&		Not. Obs.			&	$<$	4.65			&	24	\\
Gl  109     	&	41.064621	&	+25.523364	&	 M3.0	&	6.8	&	7.5	&	--	&	--	&	$<$	3.00			&		29.77	$\pm$	3.33	&	24	\\
GJ 3193 B   	&	45.464123	&	-16.593364	&	 M3.0	&	7.3	&	9.4	&	VB	&	7.1	&		25.53	$\pm$	1.32	&		95.91	$\pm$	1.60	&	24, 25	\\
GJ 1065     	&	57.684670	&	-06.094440	&	 M3.5	&	8.6	&	9.5	&	--	&	--	&	$<$	2.00			&		5.89	$\pm$	1.16	&	24	\\
Gl  176     	&	70.732396	&	+18.958168	&	 M2.0	&	6.5	&	9.3	&	--	&	--	&		10.68	$\pm$	4.09	&		126.10	$\pm$	5.13	&	24	\\
GJ 3325     	&	75.833690	&	-17.373539	&	 M3.0	&	7.8	&	9.2	&	--	&	--	&	$<$	3.00			&		10.81	$\pm$	2.42	&	24	\\
Gl  191     	&	77.919088	&	-45.018414	&	 M1.0	&	5.8	&	3.9	&	--	&	--	&	$<$	5.00			&	$<$	2.00			&	24	\\
GJ 3378     	&	90.296104	&	+59.597450	&	 M3.5	&	7.5	&	7.9	&	--	&	--	&	$<$	3.00			&	$<$	2.00			&	24	\\
Gl  226     	&	92.582703	&	+82.106756	&	 M2.0	&	6.9	&	9.4	&	--	&	--	&		0.34	$\pm$	1.84	&		36.57	$\pm$	3.13	&	24	\\
Gl  257 A  	&	104.447758	&	-44.291131	&	 M3.0	&	6.9	&	8.0	&	VB	&	1.5	&		Not. Obs.			&	$<$	3.00			&	24, 25	\\
Gl  273     	&	111.852082	&	+05.225787	&	 M3.5	&	5.7	&	3.8	&	--	&	--	&		7.60	$\pm$	2.25	&		32.34	$\pm$	3.68	&	24	\\
GJ 1105     	&	119.552907	&	+41.303690	&	 M3.5	&	7.7	&	8.6	&	--	&	--	&		11.44	$\pm$	2.62	&		27.17	$\pm$	3.56	&	24	\\
GJ 2066     	&	124.033260	&	+01.302573	&	 M2.0	&	6.6	&	9.1	&	--	&	--	&	$<$	6.00			&		53.23	$\pm$	3.89	&	24	\\
GJ 3522     	&	134.734710	&	+08.473860	&	 M3.5	&	6.5	&	6.8	&	ABC (SB+C)	&	0.6	&		104.68	$\pm$	1.66	&		376.21	$\pm$	2.00	&	24, 26	\\
GJ 1125     	&	142.685764	&	+00.322657	&	 M3.5	&	7.7	&	9.7	&	--	&	--	&	$<$	3.00			&		11.92	$\pm$	2.46	&	24	\\
Gl  358     	&	144.943205	&	-41.067558	&	 M2.0	&	6.9	&	9.5	&	--	&	--	&		8.81	$\pm$	5.29	&		Not Obs.			&	24	\\
Gl  367     	&	146.124322	&	-45.776508	&	 M1.0	&	6.6	&	9.9	&	--	&	--	&		24.23	$\pm$	5.31	&		Not Obs.			&	24	\\
Gl  382     	&	153.073621	&	-03.745666	&	 M1.5	&	5.9	&	7.9	&	--	&	--	&		18.70	$\pm$	2.81	&		220.52	$\pm$	6.50	&	24	\\
Gl  393     	&	157.231462	&	+00.841006	&	 M2.0	&	6.2	&	7.1	&	--	&	--	&		10.20	$\pm$	0.81	&		101.36	$\pm$	1.73	&	24	\\
Gl  408     	&	165.017737	&	+22.832958	&	 M2.5	&	6.3	&	6.7	&	--	&	--	&		4.32	$\pm$	2.34	&		53.59	$\pm$	4.82	&	24	\\
Gl  411     	&	165.834142	&	+35.969880	&	 M2.0	&	4.2	&	2.6	&	--	&	--	&		30.10	$\pm$	5.48	&		468.13	$\pm$	11.11	&	24	\\
Gl  412  A  	&	166.369075	&	+43.526775	&	 M0.5	&	5.5	&	4.9	&	VB (+M6)	&	35.7	&		15.94	$\pm$	1.22	&		50.66	$\pm$	1.41	&	24, 25	\\
Gl  424     	&	170.020119	&	+65.846485	&	 M0.0	&	6.3	&	8.9	&	--	&	--	&	$<$	5.00			&		131.53	$\pm$	7.33	&	24	\\
Gl  433     	&	173.862278	&	-32.539971	&	 M1.5	&	6.5	&	8.9	&	--	&	--	&		Not. Obs.			&		66.23	$\pm$	6.19	&	24	\\
Gl  445     	&	176.922406	&	+78.691163	&	 M3.5	&	6.7	&	5.4	&	--	&	--	&	$<$	3.00			&		10.82	$\pm$	1.85	&	24	\\
Gl  450     	&	177.780572	&	+35.272015	&	 M1.0	&	6.4	&	8.6	&	--	&	--	&		8.43	$\pm$	3.32	&		110.79	$\pm$	7.01	&	24	\\
Gl  465     	&	186.218762	&	-18.242290	&	 M2.0	&	7.7	&	8.9	&	--	&	--	&	$<$	3.00			&	$<$	2.00			&	24	\\
Gl  480.1   	&	190.192871	&	-43.566376	&	 M3.0	&	8.2	&	7.8	&	--	&	--	&	$<$	5.00			&	$<$	2.50			&	24	\\
Gl  514     	&	202.499109	&	+10.377164	&	 M0.5	&	5.9	&	7.7	&	--	&	--	&	$<$	5.00			&		177.59	$\pm$	6.73	&	24	\\
GJ 3801     	&	205.680292	&	+33.290099	&	 M3.5	&	7.8	&	9.3	&	--	&	--	&	$<$	5.00			&		3.38	$\pm$	2.61	&	24	\\
Gl  581     	&	229.864621	&	-07.722067	&	 M3.0	&	6.7	&	6.2	&	--	&	--	&	$<$	1.00			&	$<$	1.00			&	24	\\
Gl  623     	&	246.038854	&	+48.352906	&	 M2.5	&	6.6	&	8.1	&	--	&	--	&		7.90	$\pm$	2.17	&		53.00	$\pm$	2.71	&	24	\\
Gl  625     	&	246.352597	&	+54.304104	&	 M1.5	&	6.6	&	6.5	&	--	&	--	&	$<$	3.00			&		36.78	$\pm$	2.25	&	24	\\
GJ 1207     	&	254.273790	&	-04.348890	&	 M3.5	&	8.0	&	8.7	&	--	&	--	&		32.15	$\pm$	1.74	&		105.01	$\pm$	0.41	&	24	\\
GJ 3991     	&	257.381434	&	+43.681341	&	 M3.5	&	7.4	&	7.5	&	--	&	--	&		8.23	$\pm$	2.48	&		58.86	$\pm$	4.84	&	24	\\
Gl  678.1 A 	&	262.594696	&	+05.548531	&	 M0.0	&	6.2	&	10.0	&	VB	&	16.4	&		14.12	$\pm$	3.21	&		149.02	$\pm$	3.95	&	24, 25	\\
Gl  682     	&	264.265260	&	-44.319214	&	 M3.5	&	6.5	&	5.1	&	--	&	--	&		Not. Obs.			&		13.00	$\pm$	4.44	&	24	\\
Gl  686     	&	264.472279	&	+18.591711	&	 M1.0	&	6.4	&	8.1	&	--	&	--	&	$<$	5.00			&		90.85	$\pm$	3.60	&	24	\\
Gl  694     	&	265.983178	&	+43.378610	&	 M2.5	&	6.8	&	9.5	&	--	&	--	&	$<$	5.00			&		33.23	$\pm$	3.50	&	24	\\
Gl  693     	&	266.642624	&	-57.319043	&	 M2.0	&	6.9	&	5.8	&	--	&	--	&	$<$	6.00			&		7.33	$\pm$	3.85	&	24	\\
Gl  701     	&	271.281579	&	-03.031322	&	 M1.0	&	6.2	&	7.8	&	--	&	--	&		Not. Obs.			&		98.13	$\pm$	6.67	&	24	\\
Gl  725  A  	&	280.694497	&	+59.630409	&	 M3.0	&	5.2	&	3.6	&	VB	&	13.3	&		22.86	$\pm$	3.54	&		130.05	$\pm$	6.00	&	24, 25	\\
Gl  725  B  	&	280.695694	&	+59.626763	&	 M3.5	&	5.7	&	3.5	&	VB	&	13.3	&	$<$	3.00			&		36.07	$\pm$	3.69	&	24, 25	\\
Gl  729     	&	282.455676	&	-23.836230	&	 M3.5	&	6.2	&	3.0	&	--	&	--	&		67.53	$\pm$	6.15	&		290.53	$\pm$	7.53	&	24	\\
Gl  745  A  	&	286.773180	&	+20.888046	&	 M1.5	&	7.3	&	8.5	&	VB	&	114.2	&	$<$	5.00			&		15.57	$\pm$	3.00	&	24, 25	\\
Gl  745  B  	&	286.805013	&	+20.877011	&	 M2.0	&	7.3	&	8.8	&	VB	&	114.2	&	$<$	5.00			&		16.73	$\pm$	3.12	&	24, 25	\\
Gl  793     	&	307.633520	&	+65.449558	&	 M2.5	&	6.7	&	8.0	&	--	&	--	&	$<$	5.00			&		78.55	$\pm$	5.13	&	24	\\
Gl  809     	&	313.332460	&	+62.154390	&	 M0.5	&	5.4	&	7.1	&	--	&	--	&		Not. Obs.			&		349.48	$\pm$	9.95	&	24	\\
Gl  829     	&	322.403384	&	+17.643293	&	 M3.5	&	6.2	&	6.7	&	--	&	--	&	$<$	5.00			&		40.20	$\pm$	3.15	&	24	\\
Gl  832     	&	323.391564	&	-49.009005	&	 M1.5	&	5.3	&	5.0	&	--	&	--	&		20.85	$\pm$	2.98	&		176.59	$\pm$	5.32	&	24	\\
GJ 4248     	&	330.622304	&	-37.080894	&	 M3.5	&	7.6	&	7.5	&	--	&	--	&		Not. Obs.			&		13.27	$\pm$	3.60	&	24	\\
Gl  867  AabC  	&	339.689894	&	-20.621134	&	 M1.5	&	5.7	&	8.7	&	SB+VB	&	24.5	&		234.46	$\pm$	10.69	&		1303.88	$\pm$	15.04	&	24, 27	\\
Gl  877     	&	343.939623	&	-75.458669	&	 M2.5	&	6.6	&	8.6	&	--	&	--	&		5.25	$\pm$	1.81	&		42.24	$\pm$	2.97	&	24	\\
Gl  880     	&	344.145020	&	+16.553432	&	 M1.5	&	5.4	&	6.8	&	--	&	--	&		15.23	$\pm$	3.30	&		347.39	$\pm$	8.22	&	24	\\
Gl  908     	&	357.302200	&	+02.401224	&	 M1.0	&	5.8	&	6.0	&	--	&	--	&		13.55	$\pm$	2.40	&		130.30	$\pm$	5.16	&	24	
																										
\enddata																																																																				
\tablenotetext{a}{Distances are from Hipparcos \citep{perr97}.  Those with negative signs in front are photometric distances based on \cite{bara98} models. Hyades members without individual Hipparcos measurements are assumed to have the cluster distance of 46.34 $\pm$ 0.27 pc from \cite{perr98}.}																																																																																																																								
\tablenotetext{b}{VB = visual binary and SB = spectroscopic binary.}			
\tablenotetext{c}{Binaries with separations $<$35\arcsec\ are not resolved by \galex.}																																																																																																																					
\tablenotetext{d}{References for group members and binarity, e.g. X, Y. These are: 1:  \cite{reza89} ,
2:  \cite{bran03},
3:  \cite{zuck04},
4:  \cite{ster99},
5:  \cite{greg92},
6:  \cite{torr06},
7:  \cite{webb99},
8:  \cite{shko11a},
9:  \cite{song03},
10:  \cite{reid03b},
11:  \cite{jaya06},
12:  \cite{zuck01},
13:  \cite{loop10b},
14:  \cite{shko12},
15:  \cite{krau14},
16:  \cite{mont01},
17:  \cite{zuck04b},
18:  \cite{malo13},
19:  \cite{lope06},
20:  \cite{bowl12a},
21:  \cite{amml09},
22:  \cite{perr98},
23:  \cite{guen05},
24:  \cite{lepi11},
25:  \cite{pove94},
26:  \cite{ried14},
27: \cite{pour04}.}																																																																																																																							
\tablenotetext{d}{TWA 31 and TWA 30 B are both known to be accretors from \cite{shko11a} and \cite{loop10b}, respectively.  TWA 30 B's J-band flux is also extincted, and thus is not included in the analysis.}																											
\end{deluxetable}

\begin{deluxetable}{llccrrclccccc}																				
\tabletypesize{\scriptsize}																				
\tablecaption{Power-law coefficients for $y \propto x^\beta$ \label{coefficients}}															
\tablewidth{0pt}																				
\tablehead{																				
\colhead{$x$} 	&	 \colhead{$y$} 	&	 \colhead{Subset} 	&	 \colhead{$\beta$} 	&	 \colhead{$R$}	&	 \colhead{\#} 	
}																				
\startdata																				
($F_{NUV}/F_J$)$_{exc}$	&	($F_{FUV}/F_J$)$_{exc}$	&	all	&	1.11 $\pm$ 0.04	&	0.93	&	121	\\
($F_{NUV}/F_J$)$_{exc}$	&	($F_{FUV}/F_J$)$_{exc}$	&	strong emitters only	&	1.00 $\pm$ 0.08	&	0.79	&	102	\\
$F_{NUV}/F_J$	&	NUV surface flux	&	those with parallaxes	&	1.05 $\pm$ 0.04	&	0.94	&	103	\\
$F_{FUV}/F_J$	&	FUV surface flux	&	those with parallaxes	&	0.96 $\pm$ 0.05	&	0.91	&	72	\\
$F_{X}/F_J$	&	($F_{FUV}/F_J$)$_{exc}$	&	all	&	0.73 $\pm$ 0.06	&	0.85	&	65	\\
$F_{X}/F_J$	&	($F_{NUV}/F_J$)$_{exc}$	&	all	&	0.61 $\pm$ 0.05	&	0.83	&	65	\\
$F_{X}/F_J$	&	($F_{FUV}/F_J$)$_{exc}$	&	strong emitters only	&	0.41 $\pm$ 0.13	&	0.42	&	53	\\
$F_{X}/F_J$	&	($F_{NUV}/F_J$)$_{exc}$	&	strong emitters only	&	0.35 $\pm$ 0.12	&	0.36	&	53	\\
$F_{X}/F_J$	&	($F_{FUV}/F_J$)$_{exc}$	&	weak emitters only	&	0.34 $\pm$ 0.20	&	0.49	&	12	\\
$F_{X}/F_J$	&	($F_{NUV}/F_J$)$_{exc}$	&	weak emitters only	&	0.13 $\pm$ 0.23	&	0.18	&	53	\\
Age	&	$F_{X}/F_J$	&	medians, $\gtrsim$300 Myr	&	-1.36 $\pm$ 0.32	&	-0.97	&	3	\\ 
Age	&	($F_{FUV}/F_J$)$_{exc}$	&	medians, $\gtrsim$300 Myr	&	-0.99 $\pm$ 0.19	&	-0.98	&	3	\\ 
Age	&	($F_{NUV}/F_J$)$_{exc}$	&	 medians, $\gtrsim$300 Myr	&	-0.84 $\pm$ 0.09	&	-0.99	&	3

\enddata																				
																				
\end{deluxetable}
																																	
 \clearpage

\end{document}